\documentclass[12pt]{article}
\usepackage{axodraw}
\usepackage{pstricks}
\usepackage{color}
\usepackage{cite}
\usepackage{array}
\usepackage{epsfig}
\usepackage{amssymb}
\usepackage{graphics,graphpap}
\usepackage{amssymb}
\usepackage{amsmath}
\usepackage{slashed}
\usepackage{dsfont}

\def\eps{\varepsilon}

\newcommand{\gev}{\, {\rm GeV}}
\newcommand{\mev}{\, {\rm MeV}}

\newcommand{\mt}{m_{\rm t}}
\newcommand{\mtb}{\overline{m}_{\rm t}}

\newcommand{\be}{\begin{equation}}
\newcommand{\ee}{\end{equation}}
\newcommand{\bea}{\begin{eqnarray}}
\newcommand{\eea}{\end{eqnarray}}

\newcommand{\bi}{\begin{itemize}}
\newcommand{\ei}{\end{itemize}}
\newcommand{\ord}{{\cal O}}

\newcommand{\vcb}{|V_{cb}|}
\newcommand{\vtd}{|V_{td}|}
\newcommand{\vub}{|V_{ub}|}
\newcommand{\vts}{|V_{ts}|}
\newcommand{\vus}{|V_{us}|}

\def\kpn{K^+\rightarrow\pi^+\nu\bar\nu}

\usepackage{graphicx}

\medskipamount 

\usepackage{fancyhdr}
\advance \headheight by 3.0truept       

\newlength{\textlength}
\newlength{\overlinelength}

 \def\s#1{\setbox0=\hbox{$#1$}%
   \rlap{\ifdim\wd0>.7em\kern.22\wd0\else\kern.1\wd0\fi /}#1}


\begin{document}

\begin{titlepage}
\begin{flushright}
{FLAVOUR(267104)-ERC-40}
\end{flushright}
\vskip1.2cm
\begin{center}
{\Large \bf \boldmath
Stringent Tests of Constrained Minimal Flavour Violation through $\Delta F=2$ 
Transitions}
\vskip1.0cm
{\bf
Andrzej J. Buras and 
Jennifer Girrbach}
\vskip0.3cm
TUM-IAS, Lichtenbergstr. 2a, D-85748 Garching, Germany\\
Physik Department, TUM, D-85748 Garching, Germany\\

\vskip0.51cm


\vskip0.35cm

{\large\bf Abstract\\[10pt]} \parbox[t]{\textwidth}{
New Physics contributions to $\Delta F=2$ transitions in the simplest 
extensions of the Standard Model (SM), 
the models with constrained Minimal Flavour 
Violation (CMFV), are parametrized by a single variable $S(v)$, the value of the real
box diagram function that in CMFV is bounded from below by its 
SM value $S_0(x_t)$. With already very precise experimental 
values of $\varepsilon_K$, $\Delta M_d$, $\Delta M_s$ and 
precise values of the CP-asymmetry $S_{\psi K_S}$ and of $\hat B_K$ entering 
the evaluation of  $\varepsilon_K$,
the future of CMFV in the 
$\Delta F=2$ sector depends crucially on the values of $\vcb$,  $\vub$, $\gamma$, 
$F_{B_s} \sqrt{\hat B_{B_s}}$ and $F_{B_d} \sqrt{\hat
B_{B_d}}$.  The ratio $\xi$ of the latter two non-perturbative parameters, already rather precisely determined from lattice 
calculations, allows then together with $\Delta M_s/\Delta M_d$ and 
 $S_{\psi K_S}$ to determine the range of the angle $\gamma$ 
in the unitarity triangle independently of the value of $S(v)$. 
Imposing in addition the constraints from  
$|\varepsilon_K|$ and $\Delta M_d$ allows to determine the favorite CMFV 
values of
$\vcb$, $\vub$, $F_{B_s}\sqrt{\hat B_{B_s}}$ and $F_{B_d} \sqrt{\hat B_{B_d}}$ as 
functions of $S(v)$ and $\gamma$. The $\vcb^4$ dependence of $\varepsilon_K$ allows to 
determine $\vcb$ for a given $S(v)$ and $\gamma$ with a higher precision than it is 
presently possible using  tree-level decays. The same applies to $\vub$, $\vtd$ and $\vts$ that are 
automatically  determined as functions of $S(v)$ and $\gamma$.
We derive correlations between 
$F_{B_s}\sqrt{\hat B_{B_s}}$ and $F_{B_d} \sqrt{\hat B_{B_d}}$, $\vcb$, $\vub$ and  $\gamma$ that should be tested in the coming years. Typically 
$F_{B_s}\sqrt{\hat B_{B_s}}$ and $F_{B_d} \sqrt{\hat B_{B_d}}$ have to be significantly lower than their present lattice values, while
$\vcb$ has to be 
significantly  higher than its tree-level determination. The 
region in the space of these three parameters allowed by CMFV indicates
visible problems in this class of models and hints for the presence of new sources of flavour violation and/or new local operators in $\Delta F=2$ data
that are strongly suppressed in these models.  As a byproduct we propose to
reduce the present uncertainty in the charm contribution to $\varepsilon_K$ by using the 
experimental value of $\Delta M_K$.
}

\vfill
\end{center}
\end{titlepage}

\setcounter{footnote}{0}

\newpage

\section{Introduction}
\label{sec:1}
The simplest class of extensions of the Standard Model (SM) are models 
with constrained MFV (CMFV) \cite{Buras:2000dm,Buras:2003jf,Blanke:2006ig}
 that similarly to the SM imply stringent correlations between observables in $K$, $B_d$ and $B_s$ systems, while allowing still for significant departures 
from SM expectations. In the case of $\Delta F=2$ transitions in the 
down-quark sector, that is $K^0-\bar K^0$ and $B^0_{s,d}-\bar B^0_{s,d}$ mixings, 
new physics (NP) in these models enters only through a single universal 
real one-loop  function, the 
box diagram function $S(v)$, with $v$ standing for the parameters of a given 
CMFV model. Moreover it can be shown that in this class of models $S(v)$ is bounded from below by its SM value \cite{Blanke:2006yh}:
\be
S(v)\ge S_0(x_t)\approx 2.3
\ee
with $S_0(x_t)$ given in (\ref{S0}). 

It is known that the SM experiences 
some tension in the correlation 
$S_{\psi K_S}-\varepsilon_K$  \cite{Lunghi:2008aa,Buras:2008nn,Bona:2009cj,Lenz:2010gu,Lunghi:2010gv}. This tension 
can be naturally removed in models with CMFV by enhancing the value of $S(v)$. 
However, as pointed out in \cite{Buras:2011wi,Buras:2012ts} this spoils 
the agreement of the SM with the data on $\Delta M_{s,d}$ signaling
 the  tension between $\Delta M_{s,d}$ and $\varepsilon_K$ in CMFV models.

Now the experimental data on $\varepsilon_K$, $\Delta M_d$, $\Delta M_s$ are 
already very precise. Moreover the CP-asymmetry $S_{\psi K_S}$ is also measured 
within the accuracy of a few percent.  In all CMFV models we have 
\be\label{SpKs}
S_{\psi K_S}=\sin 2\beta \qquad {\rm (CMFV)},
\ee
and consequently this implies a universal precise value of the angle 
$\beta$ in the unitarity triangle in this class of models \cite{Buras:2000dm}.

With $\vus$ determined already very precisely the size of the tensions mentioned above depends primary on the values of 
\be\label{par}
\vcb, \quad F_{B_s}\sqrt{\hat B_{B_s}},\quad  F_{B_d} \sqrt{\hat B_{B_d}}, \quad
\hat B_K, \quad \gamma, \quad S(v),
\ee
with $\vub$ 
being then a derived quantity from the unitarity of the CKM matrix. The present 
status of these quantities can be found in Table~\ref{tab:input}. In the 
case of $B_{s,d}$ mesons we refer also to \cite{Davies:2012qf,Gamiz:2013waa}.

Now the lattice QCD calculations made an impressive progress in
the determination of $\hat B_K$ which enters the evaluation of
$\varepsilon_K$ \cite{Aoki:2010pe,Bae:2010ki,Constantinou:2010qv,Colangelo:2010et,Bailey:2012bh,Durr:2011ap}.  The most recent world 
average $\hat B_K=0.767\pm 0.010$ \cite{Laiho:2009eu}
 is very close to its large $N$ 
value $\hat B_K=0.75$ \cite{Gaiser:1980gx,Buras:1985yx}. Moreover the analyses in 
\cite{Bardeen:1987vg} and in particular in \cite{Gerard:2010jt} show that $\hat B_K$ cannot 
be larger than its large $N$ value. Therefore we expect that the final lattice
value of  $\hat B_K$  will be slightly lower than the present one. However, as the lattice 
precision is already very high, we expect that the final result will be 
very close to $0.75$ and it is a very good approximation to set simply
\be
\hat B_K=0.75~.
\ee
Indeed compared to the present uncertainty from $\vcb$ in $\varepsilon_K$ 
proceeding in this manner is fully justified.

We are then left with five variables in (\ref{par}) and we can ask the 
question which values of them would allow to fit all $\Delta F=2$ data 
within the SM and CMFV models. While such an exercise is against the 
spirit that $\gamma$ and $\vcb$ should be determined from tree-level decays, while 
$F_{B_s}\sqrt{\hat B_{B_s}}$ and $F_{B_d} \sqrt{\hat B_{B_d}}$ from lattice calculations independently of  NP contributions, in a concrete framework like 
the SM and CMFV such an exercise makes sense. In fact the determinations 
of some of the non-perturbative parameters from FCNC data was already 
performed within the SM 
with different goals in the analyses of the UTfit collaboration \cite{Bona:2007vi}. See also \cite{Lunghi:2010gv}.

The spirit of the present paper differs also from our recent studies of NP  models, in particular in
\cite{Buras:2012ts,Buras:2012sd,Buras:2012dp,Buras:2012jb,Buras:2013td,Buras:2013uqa,Buras:2013rqa}, where the emphasize has been put on
correlations 
between observables, still to be measured, rather than the correlations 
between parameters of a given NP model. However, due to the shutdown of the 
LHC  for two years
not much progress in testing these correlations is expected soon. On the other hand the coming years can be considered as lattice 
precision era, which will allow to study NP with high precision through 
$\Delta F=2$ observables for which the data are already very precise. 
In the field of FCNC processes this is a unique situation until now.

In order to reach our goal we need still one input. Ideally it should 
be the angle $\gamma$ in the unitarity triangle determined in tree-level 
decays and therefore independently of $S(v)$.  For this reason we 
present our results as functions of $\gamma$. However, as $\gamma$ is 
presently not well known from tree-level decays, we will use the range for
 the ratio
\be\label{xi}
\xi=\frac{F_{B_s}\sqrt{\hat B_{B_s}}}{F_{B_d}\sqrt{\hat B_{B_d}}}
\ee
from lattice calculations to find out which range for $\gamma$ is 
consistent with experimental value of the ratio 
$\Delta M_s/\Delta M_d$  in the full class of CMFV 
models independently of $S(v)$.

In principle, also $\vcb$ could be used as input, but the very strong 
dependence of $\varepsilon_K$ on $\vcb$ as $\vcb^4$ invites us in the 
presence of an accurate value of $\hat B_K$ to determine $\vcb$ from 
$\varepsilon_K$\footnote{The idea of determining $\vcb$ from $\varepsilon_K$ is not 
new. It was suggested in \cite{Kettell:2002ep} with the goal to 
predict  an accurate value of $\mathcal{B}(\kpn)$ within the SM that 
similarly to $\varepsilon_K$ suffers from $\vcb^4$ dependence.
But at that time 
the precision on $\hat B_K$ was 
insufficient so that this strategy could not be used efficiently.}.
 As already stated above we determine here $\vcb$ 
to fit FCNC data and this makes sense in this concrete class of models. 
Comparison 
with tree-level results for $\vcb$ will offer still another test of CMFV.

Our paper is organized as follows. In Section~\ref{sec:2} we recall the 
formulae used in our analysis. In Section~\ref{sec:3} we calculate
\be\label{par2}
\vcb, \quad F_{B_s}\sqrt{\hat B_{B_s}},\quad  F_{B_d} \sqrt{\hat B_{B_d}}, \quad
\vub, \quad \vtd, \quad \vts, \quad \mathcal{B}(B^+\to\tau^+\nu_\tau)
\ee
in CMFV models
as functions of $S(v)$ and $\gamma$. Among  these models a prominent role is played by 
 the SM.   These results will be compared in the 
future with more precise values of these quantities but already now 
one can see that CMFV models will have hard time to give a satisfactory 
description of them unless very significant differences from the present 
values of $\vcb$, $F_{B_s}\sqrt{\hat B_{B_s}}$ and $F_{B_d} \sqrt{\hat B_{B_d}}$        will be found. In Section~\ref{sec:etacc} we analyze the uncertainty 
in the determination of $\vcb$ induced by the large uncertainty in 
the QCD factor $\eta_{cc}$ in the charm contribution to $\varepsilon_K$. 
We demonstrate how this uncertainty can be reduced by a factor of three 
by using the experimental value of $\Delta M_K$ and the large $N$ estimate 
of long distance contributions to the latter.
In Section~\ref{sec:5} we address simplest models outside the CMFV framework, 
in particular the MFV models at large, where new operators and flavour blind 
phases could be present. We also comment on models with $U(2)^3$ flavour
symmetry. In Section~\ref{sec:6} we summarize briefly our results.

\section{Compendium of Basic Formulae}\label{sec:2}
\boldmath
\subsection{$\Delta F=2$ Observables}
\unboldmath
Here we collect the formulae we used in our calculations. For the mass differences in the $B^0_{s,d}-\bar B^0_{s,d}$ systems we have

\begin{equation}\label{DMS}
\Delta M_{s}=
19.1/{\rm ps}\cdot\left[ 
\frac{\sqrt{\hat B_{B_s}}F_{B_s}}{279\mev}\right]^2
\left[\frac{S(v)}{2.31}\right]
\left[\frac{\vts}{0.040} \right]^2
\left[\frac{\eta_B}{0.55}\right] \,,
\end{equation}

\begin{equation}\label{DMD}
\Delta M_d=
0.55/{\rm ps}\cdot\left[ 
\frac{\sqrt{\hat B_{B_d}}F_{B_d}}{226\mev}\right]^2
\left[\frac{S(v)}{2.31}\right]
\left[\frac{\vtd}{8.5\cdot10^{-3}} \right]^2 
\left[\frac{\eta_B}{0.55}\right]. 
\end{equation}
The value $2.31$ in the normalization of $S(v)$ is its present SM value for 
$\mtb(\mt)=163\gev$:
\be\label{S0}
S_0(x_t)  = \frac{4x_t - 11 x_t^2 + x_t^3}{4(1-x_t)^2}-\frac{3 x_t^2\log x_t}{2
(1-x_t)^3}= 2.31 \left[\frac{\mtb(\mt)}{163\gev}\right]^{1.52}~.
\ee

Concerning $\vtd$ and $\vts$, we have first 
\be
\vtd=\vus\vcb R_t
\ee
with $R_t$ being one of the length of the unitarity triangle which 
generally in CMFV can be determined independently of $S(v)$ \cite{Blanke:2006ig}:
\be\label{RT} 
R_t=\eta_R\frac{\xi}{\vus}\sqrt{\frac{\Delta M_d}{\Delta M_s}}
    \sqrt{\frac{m_{B_s}}{m_{B_d}}}, \quad \eta_R=1 -\vus\xi\sqrt{\frac{\Delta M_d}{\Delta M_s}}\sqrt{\frac{m_{B_s}}{m_{B_d}}}\cos\beta+\frac{\lambda^2}{2}+\ord(\lambda^4).
\ee
Here $\xi$ is defined in (\ref{xi})
and $\beta$ is determined also universally through (\ref{SpKs}). We also have 
\be\label{vts}
\vts=\eta_R\vcb~
\ee
with $\vcb$ determined from $\varepsilon_K$ as discussed below.

Now the departure of $\eta_R$ from unity is very small and can be calculated by using 
present lattice input for $\xi$ and the central value of $\beta$. This gives 
\be\label{RTxi}
R_t= 0.743~ \xi, \qquad \eta_R=0.9812~.
\ee
For the numerics in Section~\ref{sec:3} we however use the relation~(\ref{RT})
without  the lattice input, except for Eq.~(\ref{vcb1}) below where we use (\ref{RTxi}).

The next steps depend on whether $\xi$ or $\gamma$ is used as an input. 
If $\xi$ is used, (\ref{RTxi}) allows to determine $R_t$ and using the 
relations
\be\label{VUBG}
 R_b=(1-\frac{\lambda^2}{2})\frac{1}{\lambda}
\left| \frac{V_{ub}}{V_{cb}} \right|=
\sqrt{1+R_t^2-2 R_t\cos\beta},\qquad
\cot\gamma=\frac{1-R_t\cos\beta}{R_t\sin\beta}~.
\ee
allows to determine  $\vub$ and $\gamma$.

On the other hand if $\gamma$ from tree level decays is used as an input, 
one could determine $R_t$ and $R_b$ without 
using the lattice value of $\xi$ \cite{Buras:2002yj}:
\be\label{BPS}
R_t=\frac{\sin\gamma}{\sin(\beta+\gamma)}, \qquad R_b=\frac{\sin\beta}{\sin(\beta+\gamma)}.
\ee
In turn $\xi$ could be determined by using (\ref{RT}) and the comparison 
with its lattice value would be another test of CMFV.

Finally, the CP-violating parameter $\varepsilon_K$ is given by
\be
\varepsilon_K=\frac{\kappa_\eps e^{i\varphi_\eps}}{\sqrt{2}(\Delta M_K)_\text{exp}}\left[\Im\left(M_{12}^K\right)\right]\,,
\label{eq:3.35}
\ee
where $\varphi_\eps = (43.51\pm0.05)^\circ$ and $\kappa_\eps=0.94\pm0.02$ \cite{Buras:2008nn,Buras:2010pza} takes into account 
that $\varphi_\eps\ne \tfrac{\pi}{4}$ and includes long distance effects in $\Im( \Gamma_{12})$ and $\Im (M_{12}^K)$. Moreover ($\lambda_i
=V_{is}^*V_{id}$)
\be\label{eq:3.4}
\left(M_{12}^K\right)^*=\frac{G_F^2}{12\pi^2}F_K^2\hat
B_K m_K M_{W}^2\left[
\lambda_c^{2}\eta_{cc}S_0(x_c)+\lambda_t^{2}\eta_{tt} S(v)
+2\lambda_c\lambda_t\eta_{ct}S_0(x_c,x_t)
\right],
\ee
where the QCD factors $\eta_{ij}$   are given in Table~\ref{tab:input}
and $S_0(x_c,x_t)$ can be found in  \cite{Blanke:2011ry}. 
In writing (\ref{eq:3.4}) we make a very plausible assumption that in models 
with CMFV, NP enters only through $S(v)$. In any case the contributions involving charm quark are subleading. 

We can then expose the main parametric dependence of $\varepsilon_K$ within 
the CMFV models as follows \cite{Buras:2008nn}:
\be
|\varepsilon_K|=\kappa_\varepsilon C_\varepsilon\hat B_K\vcb^2\lambda^2\bar\eta
\left(\vcb^2(1-\bar\varrho)\eta_{tt}S(v)+\eta_{ct}S_0(x_c,x_t)-\eta_{cc}x_c\right),
\ee
where $\bar\eta$ and $\bar\varrho$ are the known variables related to the 
unitarity triangle and
\be
C_\varepsilon =\frac{G_F^2 F_K^2 m_K M_W^2}{6\sqrt{2}\pi^2(\Delta M_K)_\text{exp}}.
\ee

As in $\varepsilon_K$ the function $S(v)$ is 
proportional to $\vcb^4$ and not $\vcb^2$ as in $\Delta M_{s,d}$ it is 
$\varepsilon_K$ which plays the crucial role in fixing the favoured 
value of $\vcb$. The prime roles of $\Delta M_{s,d}$ is to determine 
$R_t$ through (\ref{RT}) and then 
 $F_{B_s}\sqrt{\hat B_{B_s}}$ and $F_{B_d} \sqrt{\hat B_{B_d}}$ through 
(\ref{DMS}) and (\ref{DMD}), respectively.

Finally, eliminating $S(v)$ from all these expressions in terms of other 
variables we find the basic formula expressing the correlation between 
various quantities discussed in our paper
\be\label{RobertB}
S_{\psi K_S}=\sin 2\beta=\frac{1}{b\Delta M_d} \left[\frac{|\varepsilon_K|}{\vcb^2\hat B_K}
-a\right],
\ee
where
\be
a=r_\varepsilon R_t\sin\beta\left[\eta_{ct} S_0(x_t,x_c)-\eta_{cc} x_c\right], \quad
b=\frac{\eta_{tt}}{\eta_B}\frac{r_\varepsilon}{2r_d\vus^2}
\frac{1}{F_{B_d}^2\hat B_{B_d}},
\ee
with 
\be
r_\varepsilon=\kappa_\varepsilon \vus^2C_\varepsilon, \quad r_d=\frac{G_F^2}{6\pi^2}M_W^2 m_{B_d}
\ee
and $R_t$ given in (\ref{RT}).

In the derivation of (\ref{RobertB}) the following relations are useful

\be
\sin 2\beta=2 \frac{\bar\eta(1-\bar\varrho)}{R^2_t}, \quad \bar\eta=R_t\sin \beta~.
\ee

\subsection{ $\mathbf{B^+\to \tau^+\nu}$ in CMFV}
It will also be interesting to consider the tree-level decay $B^+\to \tau^+\nu_\tau$. 
In the SM $B^+\to \tau^+\nu_\tau$ is
mediated by the $W^\pm$  exchange with the resulting branching ratio  given by
\begin{equation} \label{eq:Btaunu}
\mathcal{B}(B^+ \to \tau^+ \nu_\tau)_{\rm SM} = \frac{G_F^2 m_{B^+} m_\tau^2}{8\pi} \left(1-\frac{m_\tau^2}{m^2_{B^+}} \right)^2 F_{B^+}^2
|V_{ub}|^2 \tau_{B^+} = 6.05~ \vub^2\left(\frac{ F_{B^+}}{185\mev}\right)^2.
\end{equation}
Evidently this result is subject to significant parametric uncertainties induced in (\ref{eq:Btaunu}) by $F_{B^+}$ and $\vub$. However, recently the 
error on $F_{B^+}$  from lattice QCD decreased significantly \cite{Dowdall:2013tga} so that the dominant uncertainty comes from $\vub$. 

To our knowledge $B^+\to\tau^+\nu_\tau$ decay has never been considered in 
CMFV. Here we would like to point out that in this class of models the 
branching ratio for this decay is enhanced (suppressed) for the same (opposite)  sign of the lepton coupling of the new charged gauge boson 
relative to the SM one.  Indeed, the only possibility to modify the SM result 
up to 
loop corrections in CMFV is through a tree-level exchange of a new charged gauge 
boson, whose flavour interactions with quarks are governed by the CKM 
matrix. In particular the operator structure is the same.

Denoting this gauge boson by $W^\prime$ and the corresponding 
gauge coupling by $\tilde{g}_2$ one has
\be
\frac{\mathcal{B}(B^+\to\tau^+\nu)}{\mathcal{B}(B^+\to\tau^+\nu)^{\rm SM}}=
\left(1+r\frac{\tilde g_2^2}{g^2_2}\frac{M_W^2}{M_{W^\prime}^2}\right)^2,
\ee
where we introduced a factor $r$ allowing a modification 
in the lepton couplings relatively to the SM ones, in particular of its sign. 
Which sign is required will be known once the data and SM prediction improve.

If $W^\prime$ with these properties is absent, the branching ratio in this 
framework is not modified with respect to the SM up to loop 
corrections that could involve new particles but are expected to be 
small. A tree-level $H^\pm$ exchange generates new operators and is outside this framework.
In order to simplify the analysis, in what follows we will neglect direct 
NP contributions to this decay and we will use the SM formula (\ref{eq:Btaunu}). In this manner NP will enter this decay indirectly through the value 
of $\vub$ determined in our $\Delta F=2$ analysis.

\begin{table}[!tbh]
\center{\begin{tabular}{|l|l|}
\hline
$G_F = 1.16637(1)\times 10^{-5}\gev^{-2}$\hfill\cite{Nakamura:2010zzi} 	&  $m_{B_d}= m_{B^+}=5279.2(2)\mev$\hfill\cite{Beringer:1900zz}\\
$M_W = 80.385(15) \gev$\hfill\cite{Nakamura:2010zzi}  								&	$m_{B_s} =
5366.8(2)\mev$\hfill\cite{Beringer:1900zz}\\
$\sin^2\theta_W = 0.23116(13)$\hfill\cite{Nakamura:2010zzi} 				& 	$F_{B_d} =
(188\pm4)\mev$\hfill \cite{Dowdall:2013tga}\\
$\alpha(M_Z) = 1/127.9$\hfill\cite{Nakamura:2010zzi}									& 	$F_{B_s} =
(225\pm3)\mev$\hfill \cite{Dowdall:2013tga}\\
$\alpha_s(M_Z)= 0.1184(7) $\hfill\cite{Nakamura:2010zzi}			&  $F_{B^+} =(185\pm3)\mev$\hfill \cite{Dowdall:2013tga}
\\\cline{1-1}
$m_u(2\gev)=(2.1\pm0.1)\mev $ 	\hfill\cite{Laiho:2009eu}						&  $\hat B_{B_d} =1.26(11)$,  $\hat
B_{B_s} =
1.33(6)$\hfill\cite{Laiho:2009eu}\\
$m_d(2\gev)=(4.73\pm0.12)\mev$	\hfill\cite{Laiho:2009eu}							& $\hat B_{B_s}/\hat B_{B_d}
= 1.05(7)$ \hfill \cite{Laiho:2009eu} \\
$m_s(2\gev)=(93.4\pm1.1) \mev$	\hfill\cite{Laiho:2009eu}				&
$F_{B_d} \sqrt{\hat
B_{B_d}} = 226(13)\mev$\hfill\cite{Laiho:2009eu} \\
$m_c(m_c) = (1.279\pm 0.013) \gev$ \hfill\cite{Chetyrkin:2009fv}					&
$F_{B_s} \sqrt{\hat B_{B_s}} =
279(13)\mev$\hfill\cite{Laiho:2009eu} \\
$m_b(m_b)=4.19^{+0.18}_{-0.06}\gev$\hfill\cite{Nakamura:2010zzi} 			& $\xi =
1.237(32)$\hfill\cite{Laiho:2009eu}
\\
$m_t(m_t) = 163(1)\gev$\hfill\cite{Laiho:2009eu,Allison:2008xk} &  $\eta_B=0.55(1)$\hfill\cite{Buras:1990fn,Urban:1997gw}  \\
$M_t=173.2\pm0.9 \gev$\hfill\cite{Aaltonen:2012ra}						&  $\Delta M_d = 0.507(4)
\,\text{ps}^{-1}$\hfill\cite{Amhis:2012bh}\\\cline{1-1}
$m_K= 497.614(24)\mev$	\hfill\cite{Nakamura:2010zzi}								&  $\Delta M_s = 17.72(4)
\,\text{ps}^{-1}$\hfill\cite{Amhis:2012bh}
\\	
$F_K = 156.1(11)\mev$\hfill\cite{Laiho:2009eu}												&
$S_{\psi K_S}= 0.679(20)$\hfill\cite{Nakamura:2010zzi}\\
$\hat B_K= 0.767(10)$\hfill\cite{Laiho:2009eu}												&
$S_{\psi\phi}= -0.01\pm 0.08$\hfill\cite{Aaij:2013oba}\\
$\kappa_\epsilon=0.94(2)$\hfill\cite{Buras:2008nn,Buras:2010pza}				& $\Delta\Gamma_s=0.116\pm 0.019$
\cite{Raven:2012fb}
\\	
$\eta_{cc}=1.87(76)$\hfill\cite{Brod:2011ty}												
	& $\tau_{B_s}= 1.503(10)\,\text{ps}$\hfill\cite{Amhis:2012bh}\\		
$\eta_{tt}=0.5765(65)$\hfill\cite{Buras:1990fn}												
& $\tau_{B_d}= 1.519(7) \,\text{ps}$\hfill\cite{Amhis:2012bh}\\
$\eta_{ct}= 0.496(47)$\hfill\cite{Brod:2010mj}												
& $\tau_{B^\pm}=(1641\pm8)\times10^{-3}\,\text{ps}$\hfill\cite{Amhis:2012bh}    \\
$\Delta M_K= 0.5292(9)\times 10^{-2} \,\text{ps}^{-1}$\hfill\cite{Nakamura:2010zzi}	&
$|V_{us}|=0.2252(9)$\hfill\cite{Amhis:2012bh}\\
$|\eps_K|= 2.228(11)\times 10^{-3}$\hfill\cite{Nakamura:2010zzi}					& $|V_{cb}|=(40.9\pm1.1)\times
10^{-3}$\hfill\cite{Beringer:1900zz}\\
  $\mathcal{B}(B\to X_s\gamma)=(3.55\pm0.24\pm0.09) \times10^{-4}$\hfill\cite{Nakamura:2010zzi}     &
$|V^\text{incl.}_{ub}|=(4.41\pm0.31)\times10^{-3}$\hfill\cite{Beringer:1900zz}\\
$\mathcal{B}(B^+\to\tau^+\nu)=(0.99\pm0.25)\times10^{-4}$\hfill\cite{Tarantino:2012mq}\	&
$|V^\text{excl.}_{ub}|=(3.23\pm0.31)\times10^{-3}$\hfill\cite{Beringer:1900zz}\\
\hline
\end{tabular}  }
\caption {\textit{Values of the experimental and theoretical
    quantities used as input parameters.}}
\label{tab:input}~\\[-2mm]\hrule
\end{table}

\section{Numerical Analysis}\label{sec:3}
Our analysis uses the inputs of  Table~\ref{tab:input} where we also
 collect the 
present values of the quantities in (\ref{par}) that we will determine.
In this context let us emphasize that while the results for weak decay 
constants from lattice are very new, the values of  
$F_{B_s} \sqrt{\hat B_{B_s}}$, 
 $F_{B_d} \sqrt{\hat B_{B_d}}$, $\hat B_{B_s}$ and $\hat B_{B_d}$ given in this 
Table are from 
 2011. But soon these results should be updated. Still it is tempting to 
 combine the new results for weak decay constants with the values of the 
$\hat B_i$ parameters in Table~\ref{tab:input} to find 
\be\label{newf}
F_{B_s} \sqrt{\hat B_{B_s}}=259 (7)\mev, \qquad F_{B_d} \sqrt{\hat B_{B_d}}=211 (10)\mev, \qquad \xi= 1.227.
\ee
We observe that $\xi$ is basically unchanged but the values of  
$F_{B_s} \sqrt{\hat B_{B_s}}$ and $F_{B_d} \sqrt{\hat B_{B_d}}$ are decreased. 
Consequently  as investigated in \cite{Buras:2011wi} also $\Delta M_s$ and $\Delta M_d$ decrease so that the $\varepsilon_K$-$\Delta M_{s,d}$ tension in CMFV identified in 
\cite{Buras:2011wi,Buras:2012ts} would be softened. But this requires 
the confirmation of the values in (\ref{newf}) 
by future lattice calculations.

 Before entering our analysis we want to find the range for 
$\gamma$ by using the range for $\xi$ in Table~\ref{tab:input} together 
with formulae (\ref{RT}) and (\ref{VUBG}). We find 
independently of $S(v)$\footnote{We included 
the uncertainty from $S_{\psi K_S}$ as well.} 
\be\label{gammarange}
\gamma =(66.6\pm 3.7)^\circ  \qquad  {\rm (CMFV)}
\ee
which having much smaller error agrees very well with  $\gamma$ 
 from tree-level decays obtained by LHCb
\be\label{gLHCb}
\gamma=(67.2\pm 12)^\circ, \qquad  {\rm (LHCb)}~.
\ee
 Similar comment applies to  the extraction of $\gamma$ from U-spin analysis of $B_s\to K^+K^-$ and $B_d\to\pi^+\pi^-$ decays
($\gamma=(68.2\pm 7.1)^\circ$) \cite{Fleischer:2010ib}.

\begin{figure}[!tb]
\centering
\includegraphics[width = 0.65\textwidth]{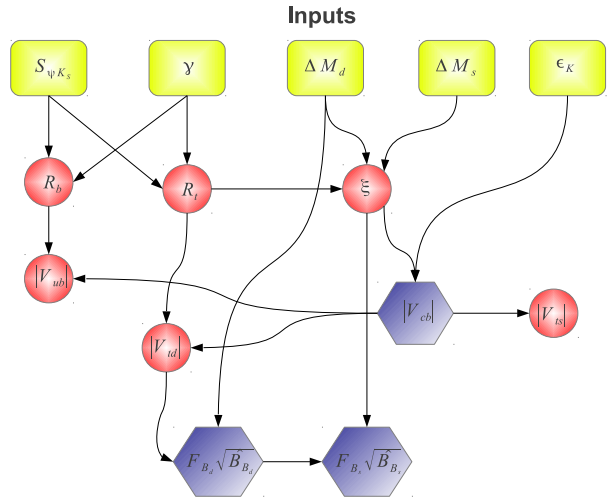}
 \caption{\it The overview of the determination of various quantities discussed 
in the text with results given in Table~\ref{tab:CMFVpred}.}\label{fig:chart}
\end{figure}

In our paper guided by the range in (\ref{gammarange}) we will 
proceed in  the following steps\footnote{The numerical values in the expressions 
in this section correspond to 
the central values of the remaining input parameters.}:

{\bf Step 1:} For a given $\gamma$ and $\beta$ from  $S_{\psi K_S}$ we will calculate $R_b$ and 
$R_t$ by means of (\ref{BPS}) and subsequently $\xi$ by using (\ref{RT}). 

{\bf Step 2:} Having $\xi$ allows with the help of 
$\epsilon_K$ to determine $\vcb$ as a function of 
 $S(v)$. Explicitly for central values of 
parameters we find  (using (\ref{RTxi}))
\be\label{vcb1}
\vcb=\frac{v(\eta_{cc},\eta_{ct})}{\sqrt{\xi S(v)}}\sqrt{\sqrt{1+h(\eta_{cc},\eta_{ct}) S(v)}-1}, \quad v(\eta_{cc},\eta_{ct}) = 0.0282,\quad 
h(\eta_{cc},\eta_{ct})=24.83~.
\ee
The second solution of the quadratic equation for $\vcb^2$ is excluded 
as it leads to a negative $\vcb^2$.
The numerical values of $h(\eta_{cc},\eta_{ct})$ and $v(\eta_{cc},\eta_{ct})$ correspond to the central 
values of $\eta_{cc}$ and $\eta_{ct}$ in Table~\ref{tab:input}. We will discuss 
the uncertainty due to these parameters in the next section.

With all this information at hand we can determine $\vub$, $\vts$ and $\vtd$ 
as functions of $S(v)$.

{\bf Step 3:}
Finally having $\vcb$ we can determine $\sqrt{\hat B_{B_d}}F_{B_d}$  from (\ref{DMD}) and 
knowing already $\xi$ from Step 1 also $\sqrt{\hat B_{B_s}}F_{B_s}$.

As the route to determine all these quantities could appear not 
transparent we show it in a chart in Fig.~\ref{fig:chart}.

\begin{figure}[!tb]
\centering
\includegraphics[width = 0.45\textwidth]{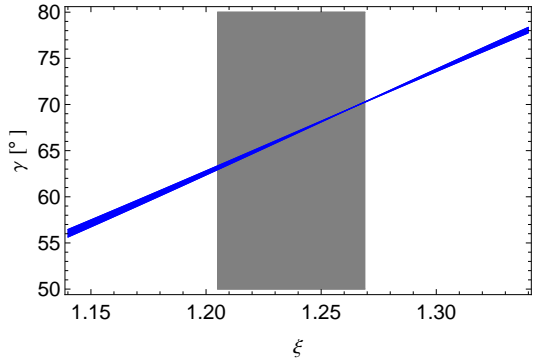}
 \caption{\it $\gamma$ versus $\xi$ for $S_{\psi K_S}\in[0.659,0.699]$.}\label{fig:gammavsxi}
\end{figure}

In Fig.~\ref{fig:gammavsxi} we show the dependence of $\xi$ on $\gamma$  varying  $S_{\psi K_S}$ in its one $\sigma$ range. We note that
the uncertainty due to $S_{\psi K_S}$ is very 
small.
This triple correlation is universal for all CMFV models including the SM and is  
central for tests of CMFV framework. Measuring $\gamma$ from tree-level 
decays precisely together with precise values of $S_{\psi K_S}$ and $\xi$ 
will be an important test for this class of models as this correlation 
is independent of $S(v)$. The range for $\gamma$ in (\ref{gammarange}) 
has been extracted from this plot by varying $\xi$ and $S_{\psi K_S}$ independently in their present one $\sigma$ ranges.

It should be emphasized that this first test is independent of $\vub$ and 
$\vcb$ and can become a precision test of CMFV models. But in order to find 
out which CMFV model, if any, is chosen by nature we have to consider 
the quantities which depend on $S(v)$. This brings us to Steps 2 and 3.

\begin{figure}[!tb]
\centering
\includegraphics[width = 0.45\textwidth]{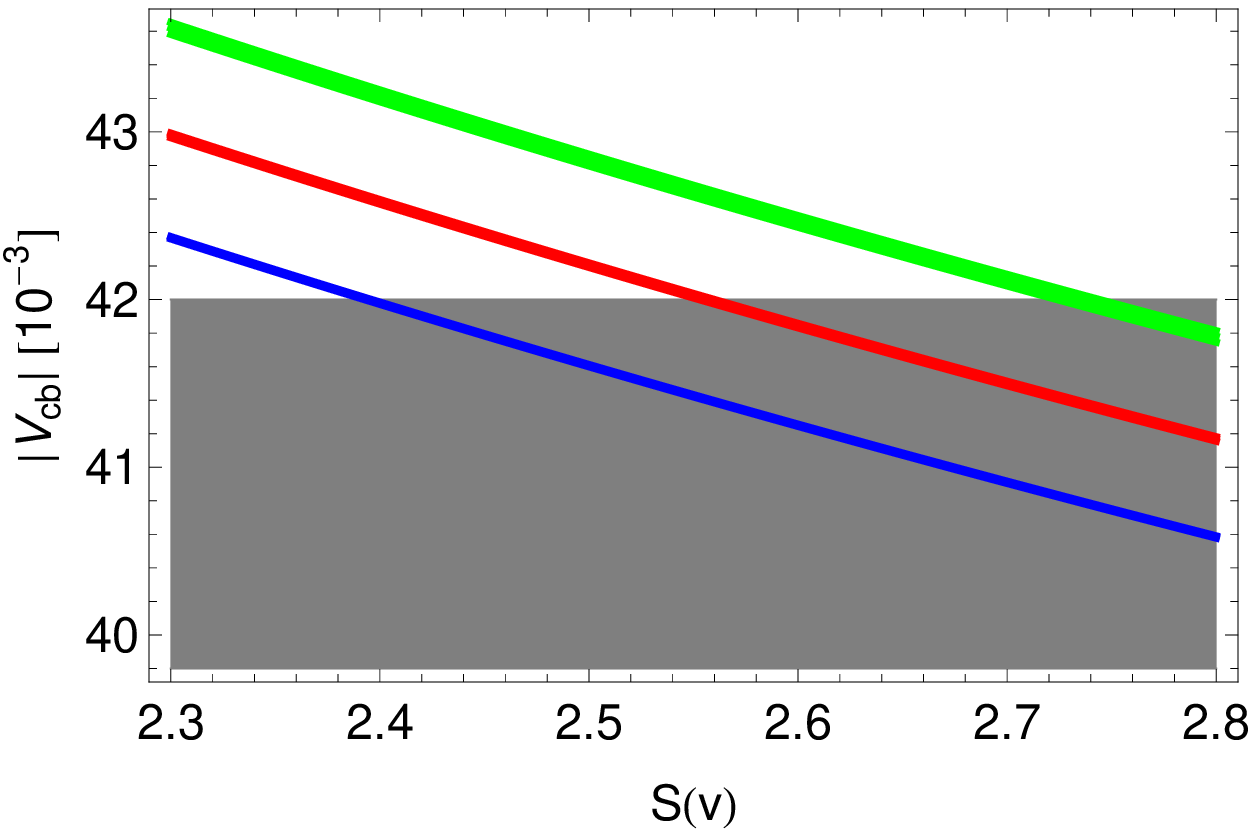} 
\includegraphics[width = 0.45\textwidth]{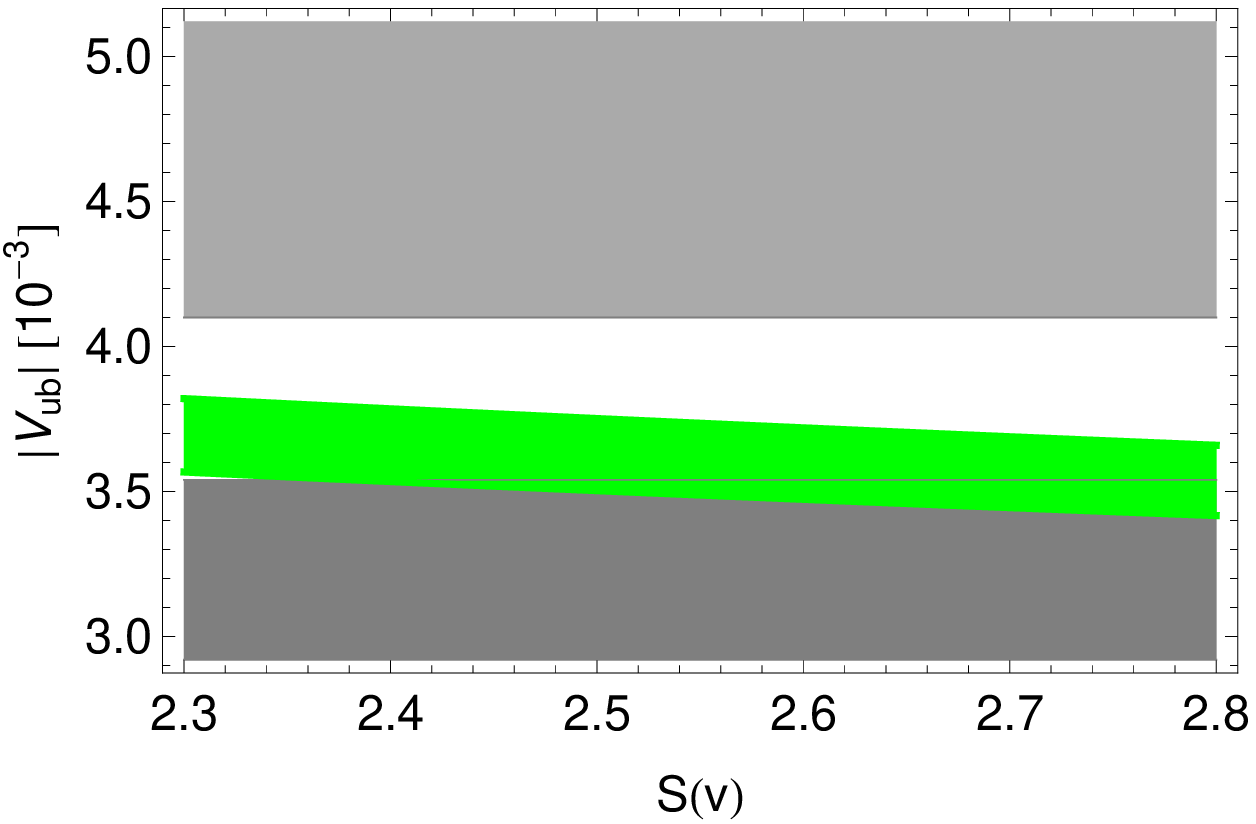}

\includegraphics[width = 0.45\textwidth]{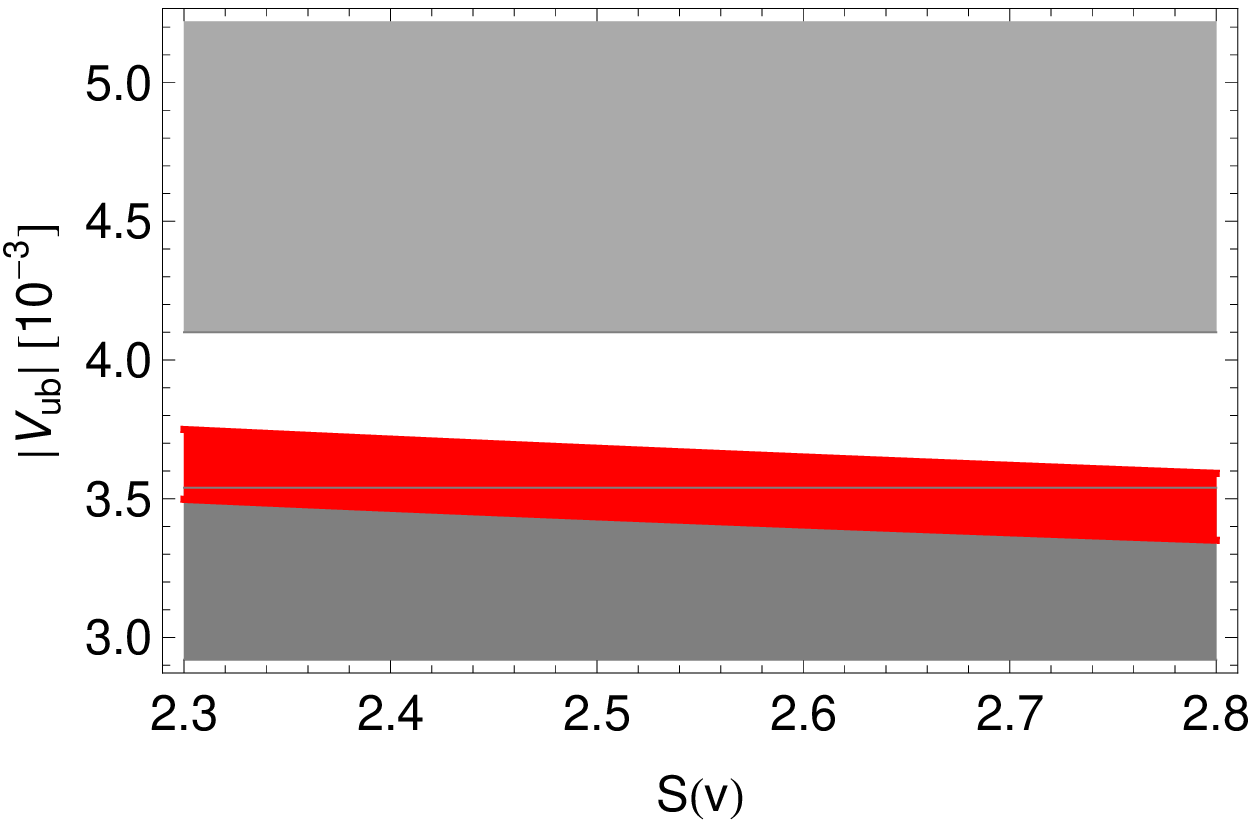}
\includegraphics[width = 0.45\textwidth]{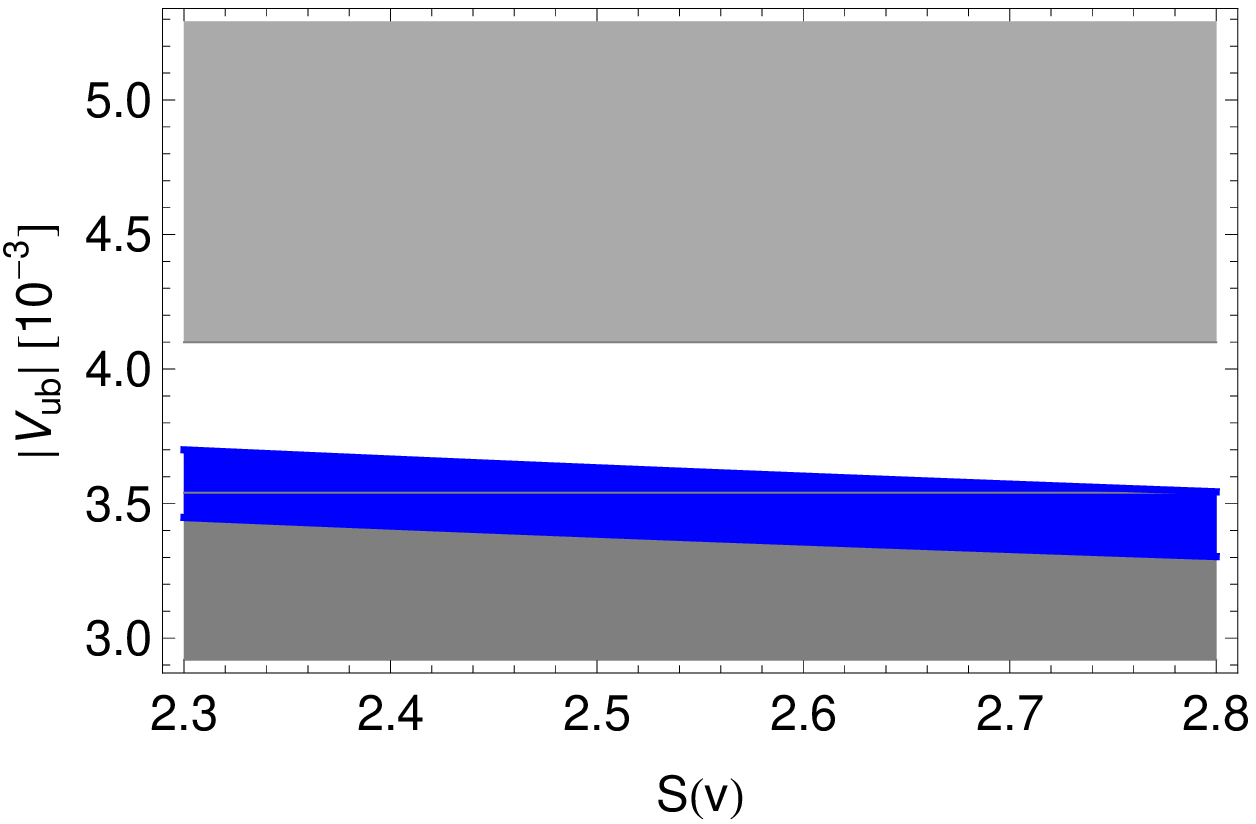}

\includegraphics[width = 0.45\textwidth]{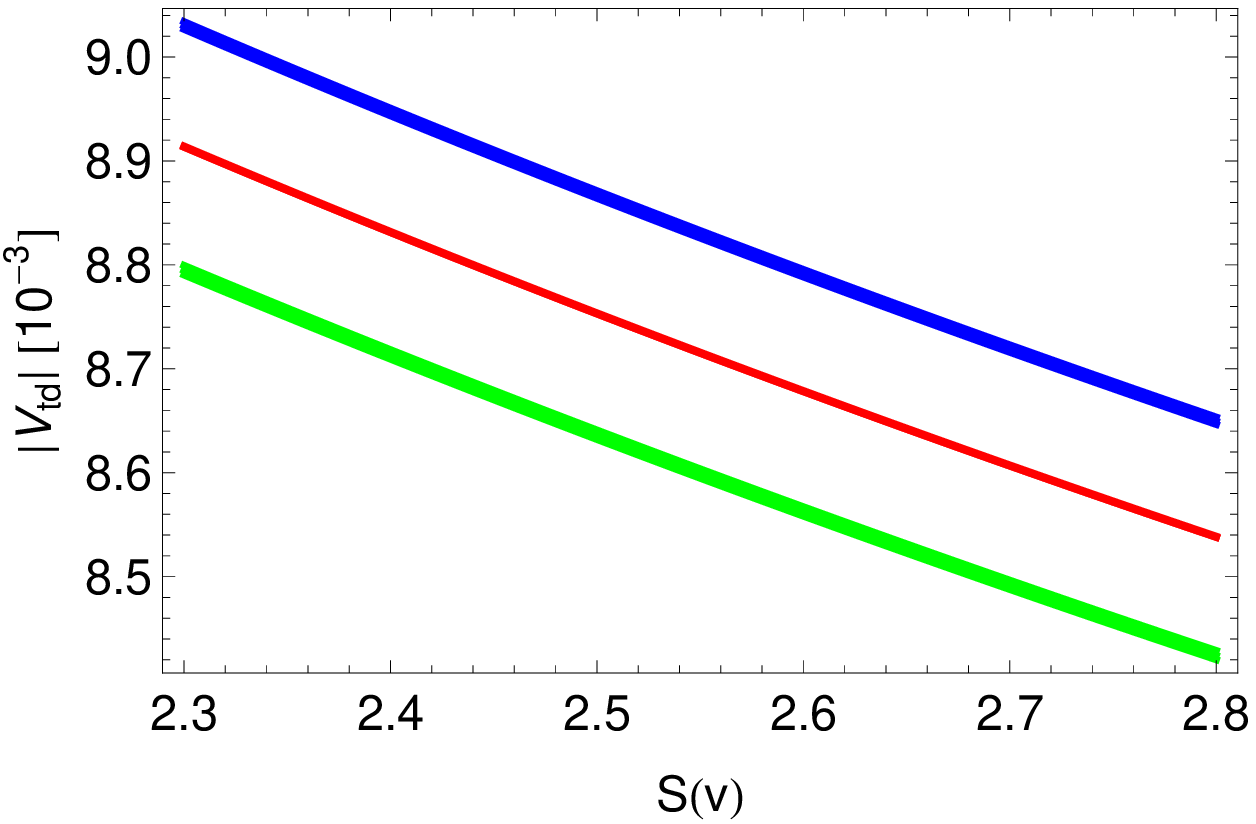}
\includegraphics[width = 0.45\textwidth]{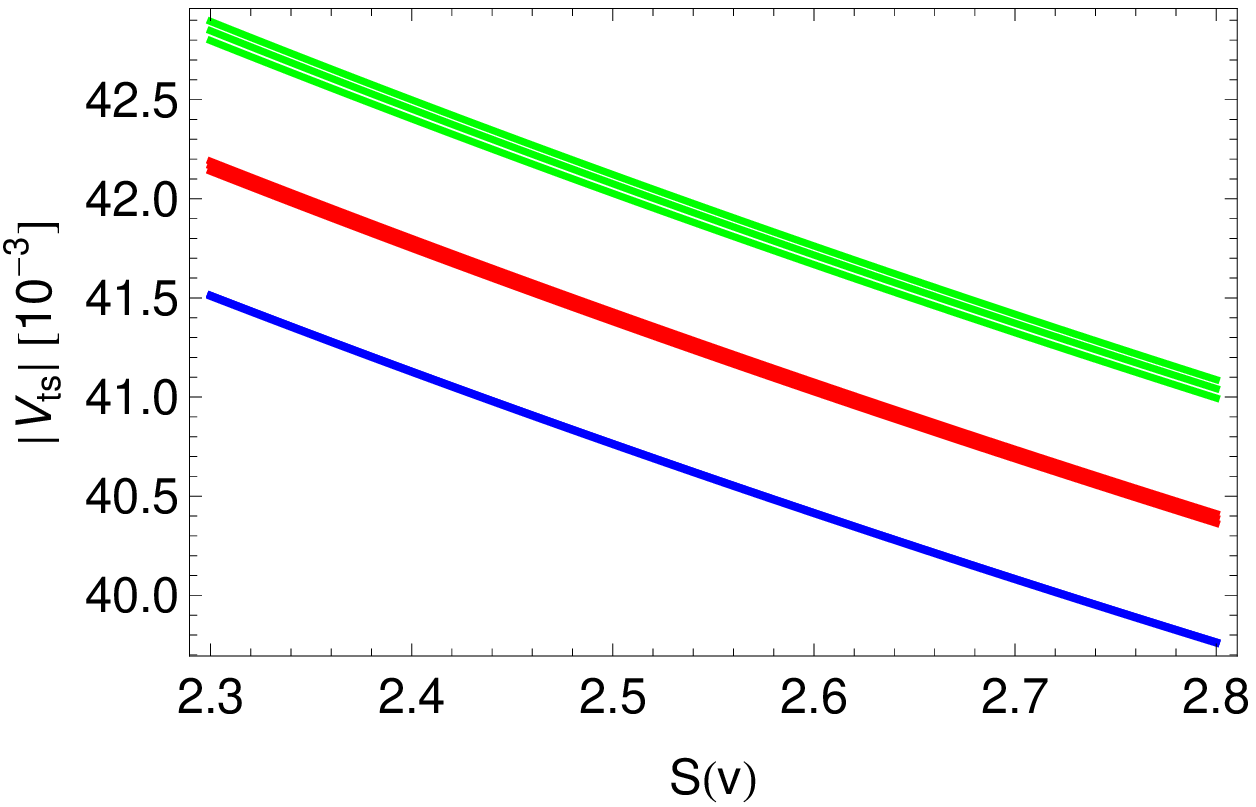}
\caption{\it CKM matrix elements versus $S(v)$ for $\gamma = 63^\circ/67^\circ/71^\circ$ (green, red,
blue). The thickness of the lines corresponds to the $1\sigma$ range of $S_{\psi K_S}\in[0.659,0.699]$.}\label{fig:VvsSv}~\\[-2mm]\hrule
\end{figure}

The $S(v)$ and $\gamma$ dependence of  all quantities of interest is shown in 
Table~\ref{tab:CMFVpred}. We show there also the SM case corresponding to  
$S(v)=S_0(x_t)=2.31$. These results are also shown 
in Fig.~\ref{fig:VvsSv}, where we plot various quantities as functions of $S(v)$ for 
different values of $\gamma$. The thickness of the lines shows the 
uncertainty in $S_{\psi K_S}$.  Evidently these 
plots are correlated with each other:
\begin{itemize}
\item
Once one of the  variables is determined, also $S(v)$ is determined 
and consequently the remaining  quantities are predicted.
\item
In a given model in which $S(v)$ is known, all  quantities considered 
are predicted.
\end{itemize}

Equivalently to Fig.~\ref{fig:VvsSv} one can find using (\ref{RobertB}) explicit universal correlations 
between various quantities in  Table~\ref{tab:CMFVpred}. We show few examples 
below. In all plots we also show the present best values of 
all quantities obtained directly from lattice or tree-level decays.

\begin{table}[!tb]
\centering
\begin{tabular}{|c|c||c|c|c|c|c|c|c|c|}
\hline
 $S(v)$  & $\gamma$ & $\vcb$ & $\vub$ & $\vtd$&  $\vts$ & $F_{B_s}\sqrt{\hat B_{B_s}}$ & 
$F_{B_d} \sqrt{\hat B_{B_d}}$ & $\xi$ &  $\mathcal{B}(B^+\to \tau^+\nu)$\\
\hline
\hline
  \parbox[0pt][1.6em][c]{0cm}{} $2.31$ & $63^\circ$ & $43.6$ & $3.69$  & $8.79$ & $42.8$ & $252.7$ &$210.0$ & $1.204$ &  $0.822$\\
 \parbox[0pt][1.6em][c]{0cm}{}$2.5$ & $63^\circ$ & $42.8$& $3.63$ & $8.64$ & $42.1$ & $247.1$ & $205.3$ &$1.204$ &  $0.794$\\
 \parbox[0pt][1.6em][c]{0cm}{}$2.7$ &$63^\circ$ & $42.1$ & $3.56$ & $8.49$ &  $41.4$  & 
$241.8$ & $200.9$ & $1.204$ &  $0.768$\\
\hline
  \parbox[0pt][1.6em][c]{0cm}{} $2.31$ & $67^\circ$ & $42.9$ & $3.62$  & $8.90$ & $42.1$ & $256.8$ &$207.2$ &$1.240$ & $0.791$\\
 \parbox[0pt][1.6em][c]{0cm}{}$2.5$ & $67^\circ$ & $42.2$& $3.56$ & $8.75$ & $41.4$ & $251.1$ & $202.6$ &$1.240$ & $0.765$\\
 \parbox[0pt][1.6em][c]{0cm}{}$2.7$ &$67^\circ$ & $41.5$ & $3.50$ & $8.61$ &  $40.7$  & 
$245.7$ & $198.3$ &$1.240$ & $0.739$\\
\hline
  \parbox[0pt][1.6em][c]{0cm}{} $2.31$ & $71^\circ$ & $42.3$ & $3.57$  & $9.02$ & $41.5$ & $260.8$ &$204.5$ &$1.276$ & $0.770$\\
 \parbox[0pt][1.6em][c]{0cm}{}$2.5$ & $71^\circ$ & $41.6$& $3.51$ & $8.87$ & $40.8$ & $255.1$ & $200.0$ &$1.276$ & $0.744$\\
 \parbox[0pt][1.6em][c]{0cm}{}$2.7$ &$71^\circ$ & $40.9$ & $3.45$ & $8.72$ &  $40.1$  & 
$249.6$ & $195.7$ &$1.276$ & $0.719$\\
 \hline
\end{tabular}
\caption{\it CMFV predictions for various quantities as functions of 
$S(v)$ and $\gamma$. The four elements of the CKM matrix are in units of $10^{-3}$,  
 $F_{B_s} \sqrt{\hat B_{B_s}}$ and $F_{B_d} \sqrt{\hat B_{B_d}}$ in units of $\mev$ and  $\mathcal{B}(B^+\to \tau^+\nu)$ in units of $10^{-4}$.
}\label{tab:CMFVpred}~\\[-2mm]\hrule
\end{table}

In order to understand the results in Table~\ref{tab:CMFVpred} let us recall 
the problems of CMFV for the present input for $\vcb$, $F_{B_s} \sqrt{\hat B_{B_s}}$ and $F_{B_d} \sqrt{\hat B_{B_d}}$ in Table~\ref{tab:input} 
\cite{Buras:2011wi,Buras:2012ts}. With the 
value of $\beta$ extracted from $S_{\psi K_S}$ and $\gamma$ from $\Delta M_s/\Delta M_d$, $\varepsilon_K$ is in the SM 
significantly smaller than its experimental 
value, while $\Delta M_{s,d}$ are significantly larger than corresponding experimental values. Therefore increasing $S(v)$, while bringing $\varepsilon_K$ 
closer to the data, shifts  $\Delta M_{s,d}$ further from their experimental 
values.  
Using for example $|V_{ub}| = 0.0034$  and $\gamma=68^\circ$ where the SM prediction and the central experimental value for $S_{\psi K_S}$
coincide, the SM
prediction for $|\varepsilon_K|^\text{SM}= 0.00186$ is below the data. Turning now to CMFV, the value of $S(v)$ at which the central
experimental value of $|\varepsilon_K|$ is reproduced turns out to be $S(v)=2.9$  \cite{Buras:2012ts} to be compared 
with $S_{\rm SM}=2.31$.
At this value of $S(v)$ with present input 
the central values of $\Delta M_{s,d}$ read
\be\label{BESTCMFV}
\Delta M_d=0.69(6)~\text{ps}^{-1},\quad  \Delta M_s=23.9(2.1)~\text{ps}^{-1}~.
\ee
They both differ from experimental values by $3\sigma$. Using another value of $\vub$ which worsen the agreement of the SM with the  experimental value for
$S_{\psi K_S}$ leads to a different value of $S(v)$ to reproduce $|\varepsilon_K|^\text{exp} = 0.002228$ and thus also~(\ref{BESTCMFV})
changes.

These problems of CMFV  can also be seen when the present central 
values of $\vcb$ and of non-perturbative parameters are inserted together 
with the data on $\Delta M_{s,d}$, $\varepsilon_K$, $\vcb$ and $\vus$ into 
(\ref{RobertB}). We find then
\begin{align}
 & S_{\psi K_S}=\sin 2\beta = 0.86\,\Rightarrow \beta = 29.8^\circ\,,\qquad R_t = 0.92
\end{align}
and thus
\begin{align}
 &R_b = 0.50\,,\qquad |V_{ub}| = 0.0047\,,\qquad \gamma = 66.4^\circ\,.
\end{align}
Evidently, the fact that $S_{\psi K_S}$ is much larger than the data 
requires the presence of new CP-violating phases. This exercise is 
equivalent to the one performed in \cite{Lunghi:2008aa}, where  $\varepsilon_K$ has been set to its experimental value but $\sin 2\beta$ was predicted.
On the other hand setting $S_{\psi K_S}$ to its experimental value  as done in \cite{Buras:2008nn} one finds that
$|\varepsilon_K|$ is significantly below the data.

\begin{figure}[!tb]
\centering
\includegraphics[width = 0.45\textwidth]{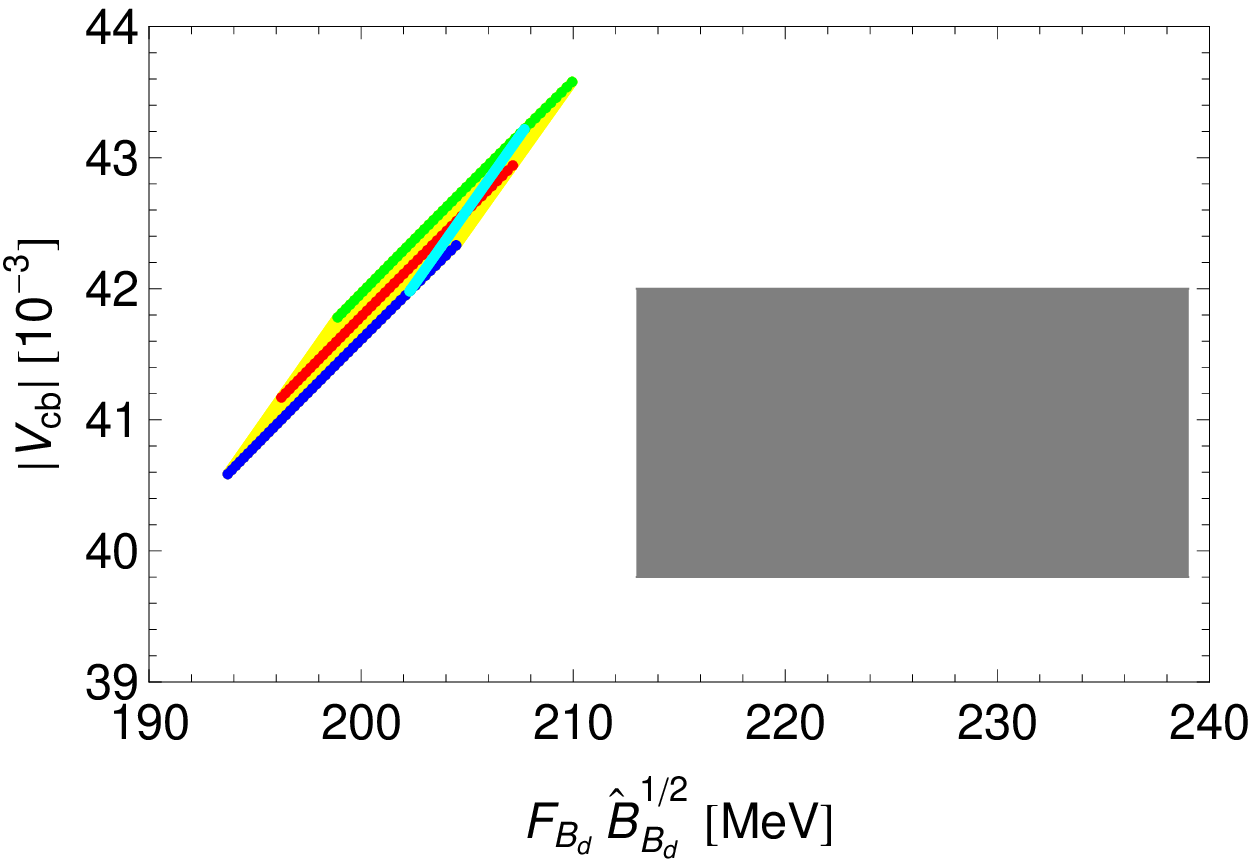}
\includegraphics[width = 0.45\textwidth]{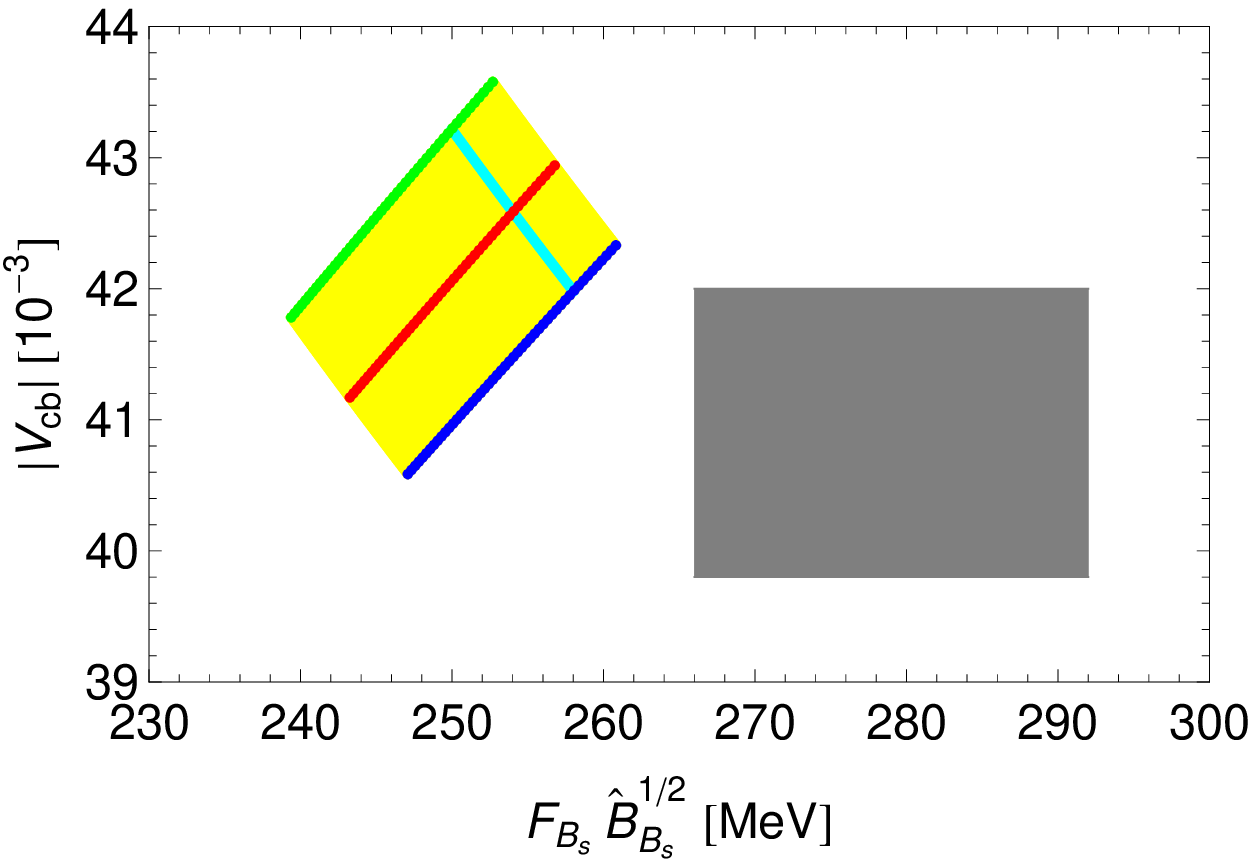}
 \caption{\it $\vcb$ versus $F_{B_d} \sqrt{\hat B_{B_d}}$ and $F_{B_s} \sqrt{\hat B_{B_s}}$ for
$S(v)\in[2.3,2.8]$ and $\gamma\in[63^\circ,71^\circ]$ (yellow), $\gamma = 63^\circ/67^\circ/71^\circ$
(green, red, blue). Cyan: fixed $S(v) = 2.4$ and $\gamma\in[63^\circ,71^\circ]$. Gray range: $1\sigma$ range of  $F_{B_d} \sqrt{\hat
B_{B_d}} = 226(13)\mev$, $F_{B_s} \sqrt{\hat B_{B_s}} =
279(13)\mev$ and $|V_{cb}|=(40.9\pm1.1)\times
10^{-3}$ (see Table~\ref{tab:input}).}\label{fig:VcbvsFB}~\\[-2mm]\hrule
\end{figure}

Thus the only hope for CMFV is that the input on $\vcb$, $F_{B_s} \sqrt{\hat B_{B_s}}$ and $F_{B_d} \sqrt{\hat B_{B_d}}$  changes with time. In particular as 
seen in Table~\ref{tab:CMFVpred}:
\begin{itemize}
\item
The values of  $F_{B_s} \sqrt{\hat B_{B_s}}$ and $F_{B_d} \sqrt{\hat B_{B_d}}$ 
have to decrease significantly, at least as far as given in (\ref{newf}). 
But this assumes the lowest value of $S(v)$. 
\item
At this value of $S(v)$ in order to obtain  agreement of $\varepsilon_K$ 
with the data 
$\vcb$ should be larger by roughly $2\sigma$ from its present central 
tree-level value.
\item
Increasing $S(v)$ allows to lower the required value of $\vcb$ but simultaneously decreases further 
$F_{B_s} \sqrt{\hat B_{B_s}}$ and $F_{B_d} \sqrt{\hat B_{B_d}}$ making their 
values significantly lower than their present values without changing their 
ratio (see Fig.~\ref{fig:VcbvsFB}). This finding shows that the freedom in choosing $S(v)$ does not 
necessarily help in solving CMFV problems.
\item
With increasing $\gamma$ the value of $\vcb$ required by $\varepsilon_K$ can 
further be decreased. For a fixed $S(v)$ this increases $F_{B_s} \sqrt{\hat B_{B_s}}$ but decreases $F_{B_d} \sqrt{\hat B_{B_d}}$ as can be
read off from Fig.~\ref{fig:VcbvsFB}.
\item
For the full range of $S(v)$ and $\gamma$ considered in Table~\ref{tab:CMFVpred}
the values of $\vub$ and $\vtd$ remain in the following ranges:
\be
\vub= (3.57\pm 0.16)\times 10^{-3}, \qquad \vtd=(8.75\pm 0.26)\times 10^{-3}
\ee
where in the case of $\vub$ we included also the error from $\beta$, which 
is irrelevant in the case of $\vtd$. Otherwise the error in $\vub$ would 
amount to $\pm 0.12$. The value of $\vts$ follows the one of $\vcb$ but 
is by $1.9\%$ smaller than the latter. Also for $\mathcal{B}(B^+\to \tau^+\nu)$ 
a narrow range is predicted:
\be
\mathcal{B}(B^+\to \tau^+\nu)=(0.77\pm 0.07)\times 10^{-4}~,
\ee
where the present uncertainty in $F_{B^+}$ has been taken into account. 
\end{itemize}

\begin{figure}[!tb]
\centering
\includegraphics[width = 0.45\textwidth]{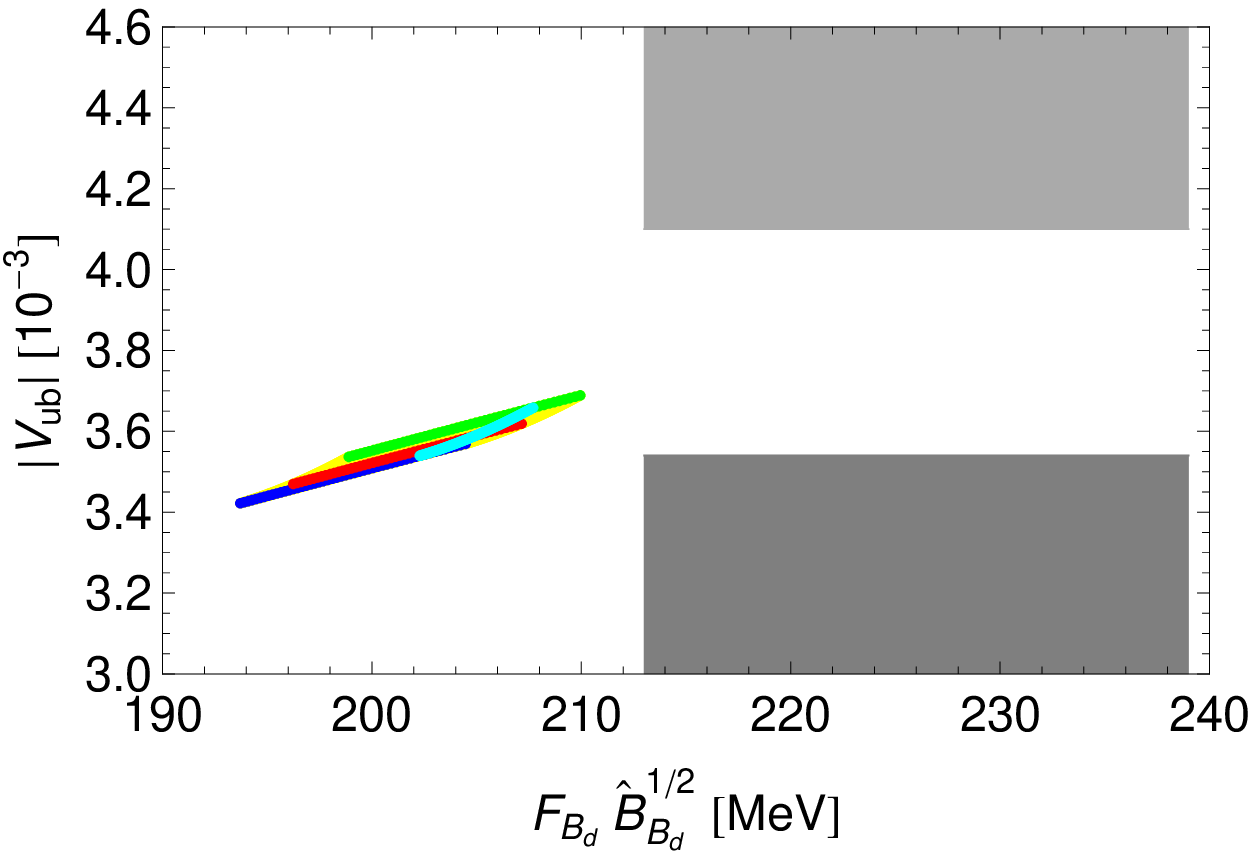}
\includegraphics[width = 0.45\textwidth]{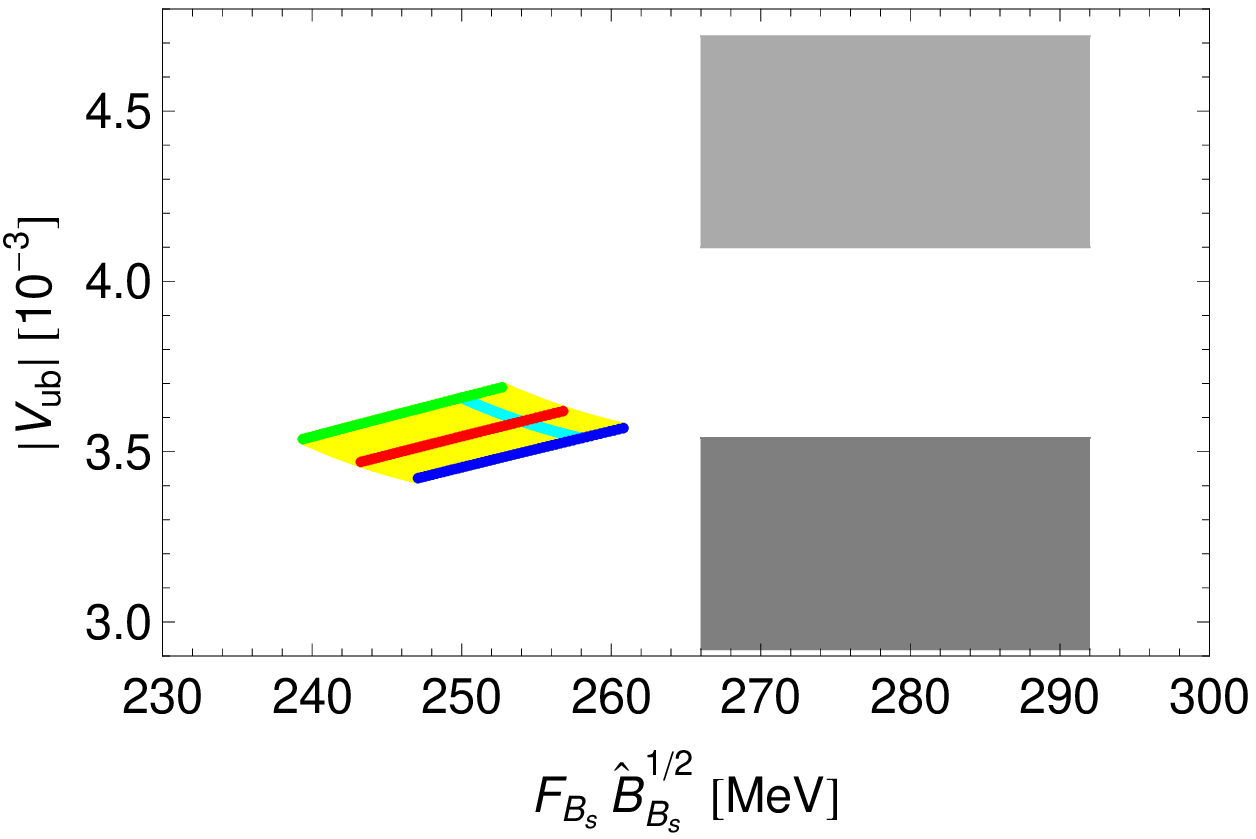}
 \caption{\it $\vub$ versus $F_{B_d} \sqrt{\hat B_{B_d}}$ and $F_{B_s} \sqrt{\hat B_{B_s}}$ for
$S(v)\in[2.3,2.8]$ and $\gamma\in[63^\circ,71^\circ]$ (yellow), $\gamma = 63^\circ/67^\circ/71^\circ$
(green, red, blue). Cyan: fixed $S(v) = 2.4$ and $\gamma\in[63^\circ,71^\circ]$. Gray range: $1\sigma$ range of  $F_{B_d} \sqrt{\hat
B_{B_d}} = 226(13)\mev$, $F_{B_s} \sqrt{\hat B_{B_s}} =
279(13)\mev$ and $|V_{ub}^\text{excl}|=(3.21\pm0.31)\times
10^{-3}$ (light gray), $|V_{ub}^\text{incl}|=(4.41\pm0.31)\times
10^{-3}$ (dark gray) (see Table~\ref{tab:input}).}\label{fig:VubvsFB}~\\[-2mm]\hrule
\end{figure}

In Fig.~\ref{fig:VcbvsFB} on the left hand side we show the correlation between $F_{B_d} \sqrt{\hat B_{B_d}}$ and $\vcb$ for different value
of $\gamma$. Analogous correlation 
between  $F_{B_s} \sqrt{\hat B_{B_s}}$ and $\vcb$ is shown on the right hand side. 
Possibly these two plots showing the allowed ranges 
for the three parameters in question in the CMFV framework are the 
most important result of our paper.
Similarly we show in Fig.~\ref{fig:VubvsFB} the same correlation but with $\vub$.

\boldmath
\section{The Uncertainty due to  $\eta_{cc}$ and $\eta_{ct}$}\label{sec:etacc}
\unboldmath
It is known that significant uncertainty  in the SM prediction for $\varepsilon_K$ comes from the value of the QCD correction $\eta_{cc}$ and to a lesser extent from $\eta_{ct}$ that 
even at the NNLO level are known only with the accuracy of $\pm 41\%$ and $\pm 9\%$, respectively \cite{Brod:2011ty,Brod:2010mj}. In our analysis this 
uncertainty enters the coefficient $h(\eta_{cc},\eta_{ct})$ of $S(v)$ and $v(\eta_{cc},\eta_{ct})$ in (\ref{vcb1}). In what follows we would like to
investigate the impact of these uncertainties 
on the determination of $\vcb$ from $\varepsilon_K$ and propose a method 
how the uncertainty in $\eta_{cc}$ could be reduced with the help of the experimental 
value of $\Delta M_K$ accompanied in particular by future lattice calculations 
of long distance effects in $\Delta M_K$.

\begin{table}[!tb]
\centering
 \begin{tabular}{|cc||ccc|}
\hline
 & & & $\eta_{cc}$  & \\
&   $h(\eta_{cc},\eta_{ct})$& 1.10 & 1.87 & 2.64\\
\hline
\hline
 & 0.451 & 18.83 & 34.93 & 85.82\\
$\eta_{ct}$ & 0.496 & 15.57 & 24.83 & 51.42\\
& 0.541 & 11.61 & 18.55 & 34.21\\
\hline
 \end{tabular}
 \begin{tabular}{|cc||ccc|}
\hline
 & & & $\eta_{cc}$  & \\
&   $v(\eta_{cc},\eta_{ct})$& 1.10 & 1.87 & 2.64\\
\hline
\hline
& 0.451 &0.0303 & 0.0259 & 0.0207\\
$\eta_{ct}$ & 0.496 &0.0323& 0.0282 & 0.0235\\
& 0.541 & 0.0341& 0.0304 & 0.0260\\
\hline
 \end{tabular}
\caption{\it Coefficient  $h(\eta_{cc},\eta_{ct})$ and $v(\eta_{cc},\eta_{ct})$ in Eq.~(\ref{vcb1}) for different values of $\eta_{cc,ct}$.
}\label{tab:heta}~\\[-2mm]\hrule
\end{table}

\begin{figure}[!tb]
 
\centering
\includegraphics[width = 0.5\textwidth]{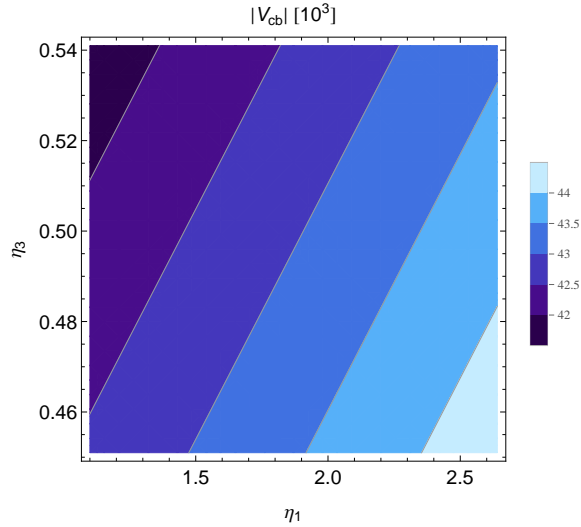}
 \caption{\it $\vcb$ as a function of $\eta_{cc}$ and $\eta_{ct}$ for fixed $\gamma = 67^\circ$ and $S(v) = S_0(x_t) = 2.31$.
}\label{fig:Vcbeta}~\\[-2mm]\hrule
\end{figure}

In Table~\ref{tab:heta} we show the values of $h(\eta_{cc},\eta_{ct})$ corresponding 
to the range of these QCD corrections calculated in \cite{Brod:2011ty,Brod:2010mj}.  
If we fix $\gamma = 67^\circ$ and $S(v) = S_0(x_t) = 2.31$ and vary $\eta_{cc}\in[1.10,2.64]$ and $\eta_{ct}\in[0.451,0.541]$
simultaneously then
$|V_{cb}|$ lies within the range $[41.7,44.4]\cdot 10^{-3}$ (see also Fig.~\ref{fig:Vcbeta}). 
Scanning only $\eta_{cc} (\eta_{ct})$ and fixing $\eta_{ct}(\eta_{cc})$ to its central value we find $\vcb\in[42.1,43.9]\cdot 10^{-3}
(\vcb\in[42.6,43.4]\cdot 10^{-3})$.
Translated in the uncertainty in the determination of $\vcb$ we find 
an uncertainty of $\pm 2.0\%$ and $\pm 0.9\%$ due to $\eta_{cc}$ and $\eta_{ct}$, 
respectively. The uncertainty due to $\eta_{tt}$ is fully negligible. It is 
expected that an improved matching of short distance calculations of $\eta_{cc}$ 
and $\eta_{ct}$ to the lattice calculations of $\hat B_K$ can significantly 
reduce the present total theoretical error on $\varepsilon_K$ and thus 
allow a more accurate extraction of $\vcb$.

As the large uncertainty in $\eta_{cc}$ is disturbing we propose to extract this 
parameter from the experimental value of $\Delta M_K$. Assuming that in CMFV 
models $\Delta M_K$ is fully dominated by the SM contributions we 
decompose it as follows:
\be 
\Delta M_K = (\Delta M_K)_{cc}+(\Delta M_K)_{ct}+(\Delta M_K)_{tt}+(\Delta M_K)_{\rm LD},
\ee 
with the first three short distance contributions obtained from 
\be
(\Delta M_K)_{ij}=2 \Re (M_{12}^K)_{ij}
\ee
with $M_{12}^K$ given in (\ref{eq:3.4}) and $i=c,t$. For the dominant 
contribution we have
\be
(\Delta M_K)_{cc}=(\lambda-\frac{\lambda^3}{2})^2
\frac{G_F^2}{6\pi^2}F_K^2 \hat B_K m_K \eta_{cc} m_c^2(m_c)~.
\ee
We find then 
setting $m_c(m_c)$ at its central value\footnote{In principle the
determination of $\eta_{cc} x_c$ is more useful as this product should be independent of 
the scale $\mu_c$ in $m_c$ but in order to compare with the value of $\eta_{cc}$ 
in Table~\ref{tab:input} we prefer here to work with $\eta_{cc}$.}
\be\label{eta1exp}
\eta_{cc}= 2.123 \left(\frac{0.75}{\hat B_K}\right) w,
\ee
where ($k=tt,ct,{\rm LD}$)
\be
w = 1 -r_{tt} - r_{ct}- r_{\rm LD}, \qquad  
r_k=\frac{(\Delta M_K)_k}{(\Delta M_K)_{\rm exp}}.
\ee

For $r_{tt}$ and $r_{ct}$ we find  within the SM
\be
r_{tt}=0.0021, \qquad r_{ct}= 0.0030
\ee 
confirming the result in \cite{Brod:2011ty} that both corrections are below 
$1\%$ and totally negligible within the SM compared with other uncertainties.  As our analysis shows in CMFV models $r_{tt}$ could be increased by 
$30\%$ due to the increase of $S(v)$ but this would still keep these corrections  below $1\%$.
Thus our 
proposal depends crucially on the estimate of $r_{\rm LD}$.  Yet, as 
we argue below the error from this contribution is significantly smaller 
than the error from the direct NNLO calculation in \cite{Brod:2011ty}.

Basically the only results on  $r_{\rm LD}$ in QCD that are available are 
from large $N$ QCD calculations in which at low energies one uses 
a dual representation of QCD as a theory of weakly interacting mesons 
\cite{'tHooft:1973jz,'tHooft:1974hx,Witten:1979kh}. In the case of 
 $K^0-\bar K^0$ mixing and non-leptonic $K$-meson decays 
this approach, developed in \cite{Buras:1985yx,Bardeen:1986vp,Bardeen:1986uz,Bardeen:1986vz,Bardeen:1987vg}, provided already a quarter of century ago results 
which are now basically confirmed by the more sophisticated lattice calculations.  This is the case of the $\hat B_K$ parameter calculated first in 
\cite{Buras:1985yx,Bardeen:1987vg} and also the case of $\Delta I=1/2$ rule 
 \cite{Bardeen:1986vz}, where the dynamics behind this rule related predominantly to current-current operators has been identified 
and the enhancements of $\Delta I=1/2$ transitions and suppression of 
$\Delta I=3/2$ transitions have been computed reaching rough agreement with 
the data. Precisely this understanding is presently emerging from lattice 
calculations \cite{Blum:2012uk,Boyle:2012ys}. 

Motivated by the success  of this approach, its results for $r_{\rm LD}$ 
could also be approximately correct..
The leading in $N$ contribution comes from 
one-loop contributions induced by two $\Delta S=1$ transitions with 
virtual $\pi\pi$, $\pi K$ and $KK$ in the loop. One finds then 
$r_{\rm LD}$ to be {\it positive} and 
in the ballpark of $0.3$ \cite{Bijnens:1990mz,Gerard:1990dx}. The 
study of subleading corrections indicates that similarly to the case 
of $\hat B_K$ these corrections are not large and tend to cancel partly 
each other 
\cite{Donoghue:1983hi,Buras:1985yx,Gerard:2005yk,Buras:2010pza} with 
some tendency to have opposite sign to the leading term \cite{Gerard:2005yk}.
 While precise calculation of $r_{\rm LD}$   in view of these cancellations is very difficult by analytic methods, based on these studies 
we expect that $r_{\rm LD}$ is likely to be positive and in the 
ballpark of  $0.1-0.3$. 
Taking 
this estimate at face value  and using (\ref{eta1exp}) we end up with 
\be\label{etaBG}
\eta_{cc}\approx 1.70 \pm 0.21.
\ee
This is consistent with $\eta_{cc}=1.87$ used by us in the previous section. 
Moreover, the error is by a factor of 3-4 smaller than the error obtained by 
the direct NNLO calculation of $\eta_{cc}$ in \cite{Brod:2011ty}. 
Interestingly   the authors of the latter paper would find this result if 
they varied the scale $\mu_c$ in the range $1.3\le\mu_c\le 1.8\gev$ and 
not in the range $1.0-2.0\gev$.

Needless to say, we are aware of the fact that these expectations and the 
estimate in (\ref{etaBG}) require more detailed investigations and in 
particular future confirmation from lattice 
simulations. Presently no reliable result on $r_{\rm LD}$ from lattice is 
available but an important progress towards its evaluation has been 
made in \cite{Christ:2012se}. This first result seems to indicate that 
$r_{\rm LD}$ could be larger than expected by us. We are therefore looking forward to more precise evaluation of this important quantity from the lattice in order 
to see whether also in this case large $N$ approach passed another test or not.

\begin{figure}[!tb]
 
\centering
\includegraphics[width = 0.45\textwidth]{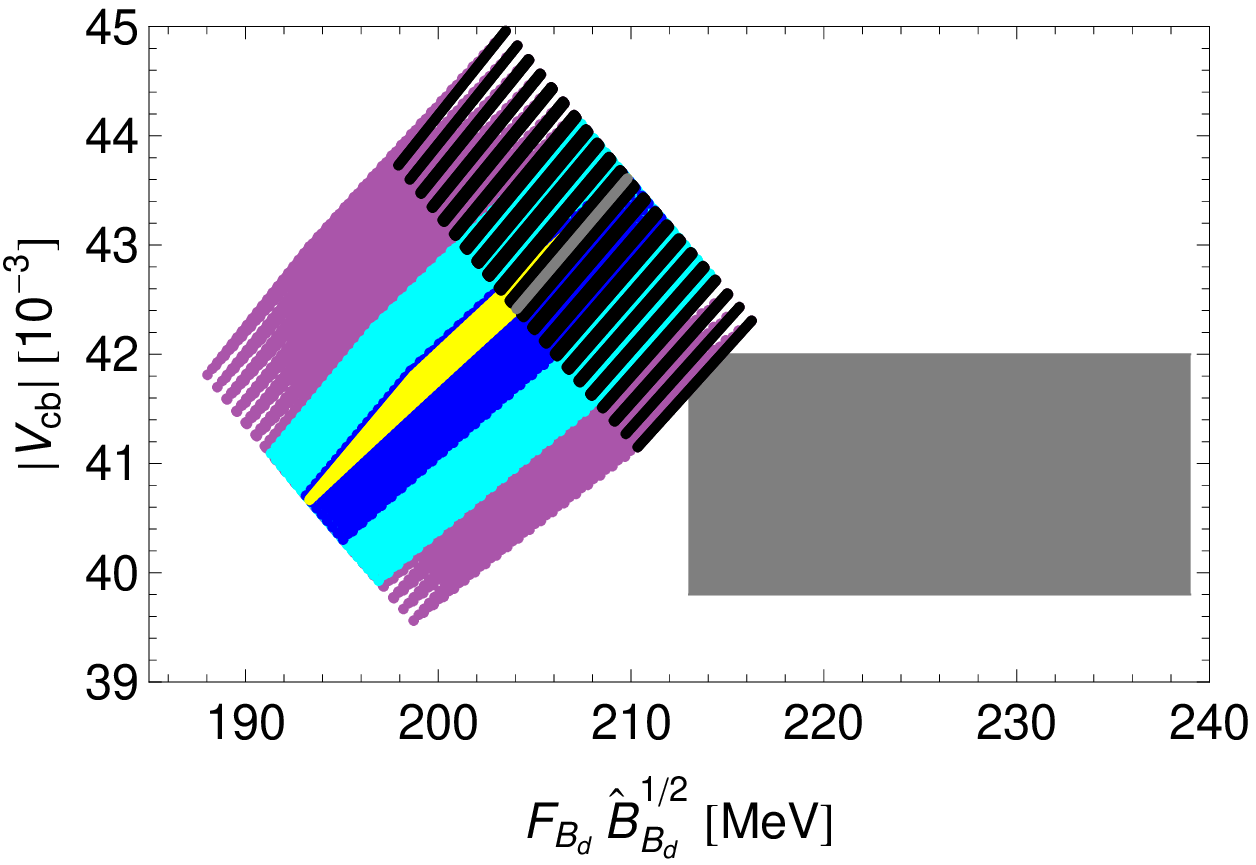}
\includegraphics[width = 0.45\textwidth]{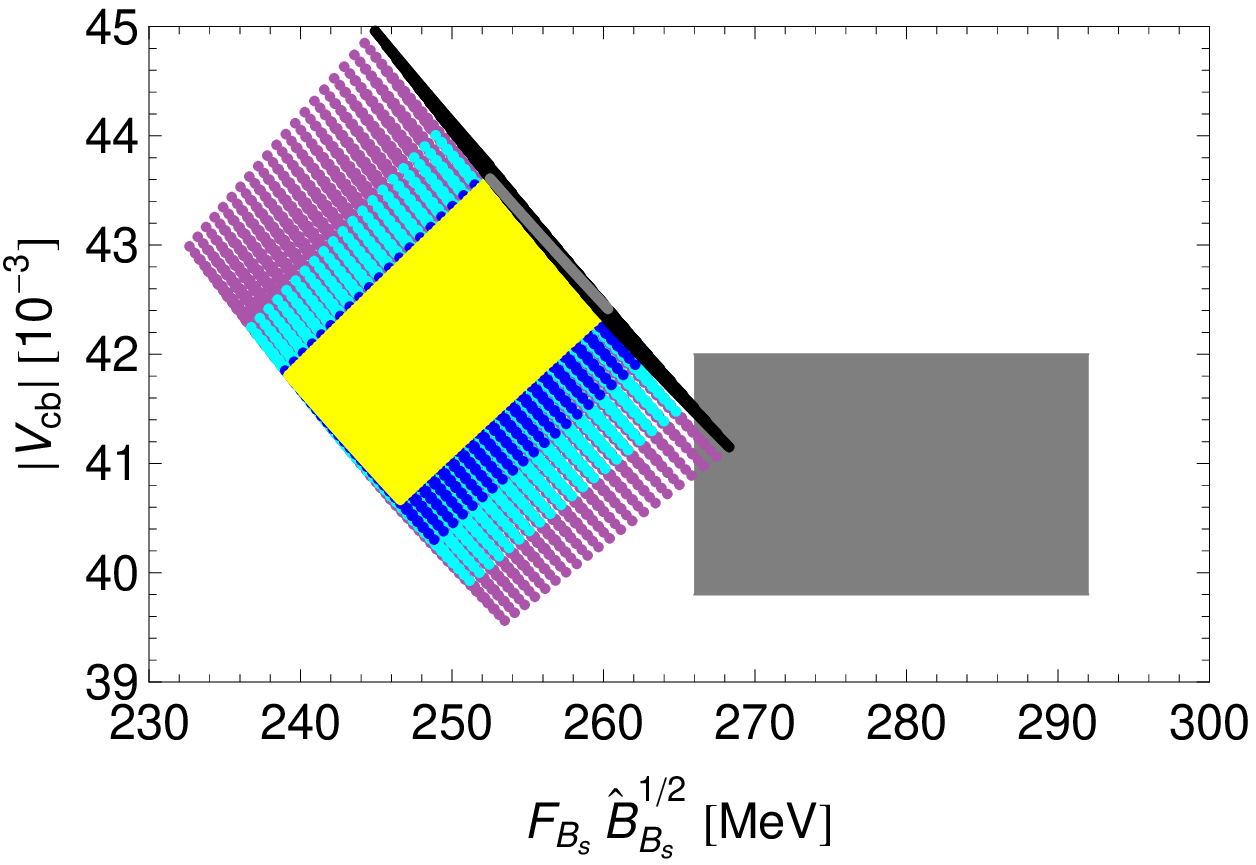}

 \caption{\it $\vcb$ versus $F_{B_d} \sqrt{\hat B_{B_d}}$ and $F_{B_s} \sqrt{\hat B_{B_s}}$ for
 $\gamma\in[63^\circ,71^\circ]$. The yellow region is the same as in Fig.~\ref{fig:VcbvsFB}: $S(v)\in[2.31,2.8]$, $\eta_{cc} =
1.87$, $\eta_{ct} = 0.496$. In the purple region we include the errors in $\eta_{cc,ct}$ as in Table~\ref{tab:heta}:  $S(v)\in[2.31,2.8]$,
$\eta_{cc} \in [1.10,2.64]$,
$\eta_{ct} \in [0.451,0.541]$. In the cyan region we use instead the reduced error of $\eta_{cc}$ as in Eq.~(\ref{etaBG}): 
$S(v)\in[2.31,2.8]$,
$\eta_{cc} \in [1.49,1.91]$,
$\eta_{ct} \in [0.451,0.541]$. In the blue region we fix $\eta_{ct}$ to its central value: $S(v)\in[2.31,2.8]$,
$\eta_{cc} \in [1.49,1.91]$,
$\eta_{ct} =0.496$. To test the SM we include the black region for fixed $S(v) = S_0(x_t) = 2.31$ and $\eta_{cc,ct}$ as in the purple
region. The gray line within the black SM region corresponds to   $\eta_{cc} =
1.87$ and $\eta_{ct} = 0.496$. The gray box is the same as in Fig.~\ref{fig:VcbvsFB}.
}\label{fig:VcbvsFBscan}~\\[-2mm]\hrule
\end{figure}

\begin{figure}[!tb]
 
\centering
\includegraphics[width = 0.45\textwidth]{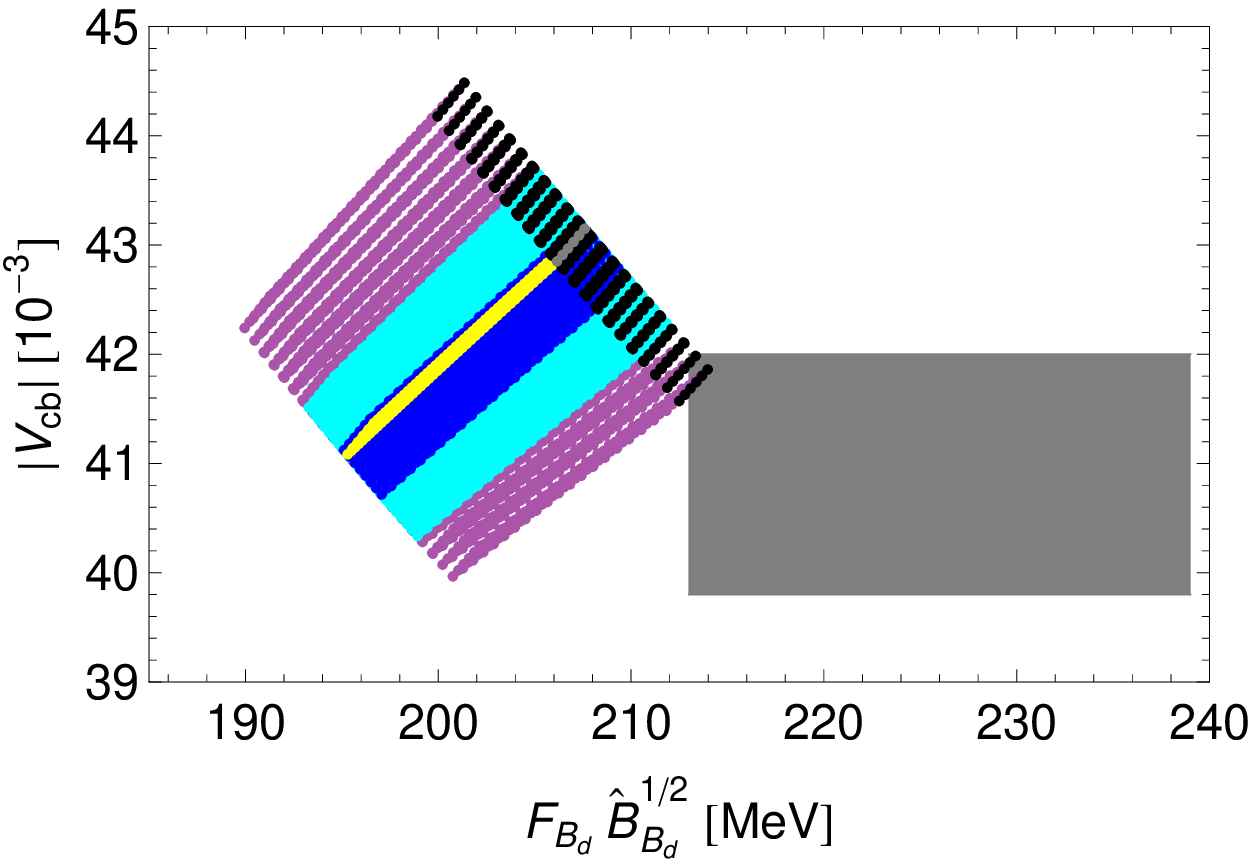}
\includegraphics[width = 0.45\textwidth]{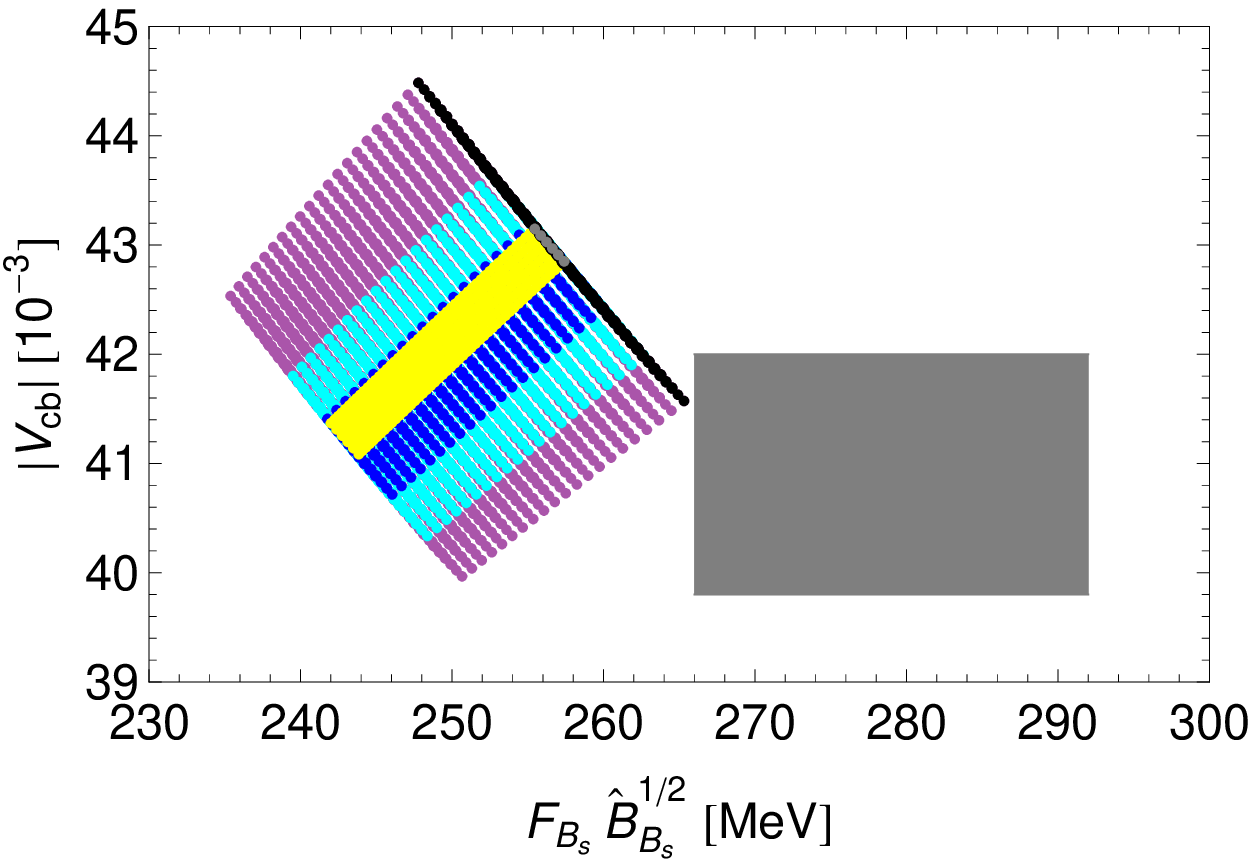}

 \caption{\it $\vcb$ versus $F_{B_d} \sqrt{\hat B_{B_d}}$ and $F_{B_s} \sqrt{\hat B_{B_s}}$ as in Fig.~\ref{fig:VcbvsFBscan} but for 
 $\gamma = (67\pm1)^\circ$. 
}\label{fig:VcbvsFBscan2}~\\[-2mm]\hrule
\end{figure}

 In Fig.~\ref{fig:VcbvsFBscan} we show the anatomy of various uncertainties  
with different ranges described in the figure caption. We observe that the 
reduced error on $\eta_{cc}$ corresponding to the cyan region 
decreased the allowed region  which with future 
lattice calculations could be decreased further. Comparing the blue 
and cyan regions we note that the reduction in the error on $\eta_{ct}$ 
would be welcomed as well. It should also be stressed that in a given 
CMFV model with fixed $S(v)$ the uncertainties are reduced further. 
This is illustrated with the black range for the case of the SM. 
 Finally an  impact on  Fig.~\ref{fig:VcbvsFBscan} will have a
precise measurement of $\gamma$ or equivalently precise lattice determination 
of 
$\xi$.  We illustrate this impact in Fig.~\ref{fig:VcbvsFBscan2} by setting in the plots of  Fig.~\ref{fig:VcbvsFBscan} $\gamma=(67\pm1)^\circ$.

\section{Going Beyond CMFV}\label{sec:5}
It is evident from our analysis and in particular from Fig.~\ref{fig:VcbvsFB}
 that in the likely case in which the central 
values of $\vcb$, $F_{B_s} \sqrt{\hat B_{B_s}}$ and $F_{B_d} \sqrt{\hat B_{B_d}}$
will only change by a few percent
but their uncertainties will be significantly reduced, 
one will have to look beyond the CMFV framework to understand the $\Delta F=2$ 
data. 
Models like Littlest Higgs model with T-parity, various Randall-Sundrum 
 scenarios and general supersymmetric models having many free parameters will be able to remove all these tensions although improved precision on 
the quantities in question would imply new constraints on the parameters of 
these models. Here we mention simpler models were the discussion can be made 
more transparent than in those more complicated models.

First in the case of MFV at large \cite{Chivukula:1987py,Hall:1990ac,D'Ambrosio:2002ex} in which new operators can contribute the situation looks a bit better.
Here the presence of left-right scalar operators in 
a 2HDM or the MSSM naturally interferes destructively with the SM contribution
 suppressing in particular 
$\Delta M_s$ \cite{Buras:2002wq}, 
but only slightly $\Delta M_d$ and having no impact on $\varepsilon_K$. Even if in these models charged Higgs and SUSY particles can enhance
$S(v)$ one 
can expect that still values of $\vcb$ will be required to be 
above the ones quoted 
in Table~\ref{tab:input} and the ones of $F_{B_s} \sqrt{\hat B_{B_s}}$ and 
$F_{B_d} \sqrt{\hat B_{B_d}}$ have to be suppressed. But the problem will 
be softer than in CMFV. On the other hand left-right operators affect the 
ratio $\Delta M_s/\Delta M_d$ which in CMFV models agrees well with data.
The solution then would be left-left scalar operators which affect 
$\Delta M_s$ and $\Delta M_d$ by the same factor and keep this ratio fixed.

The latter solution appears to be the best for 
the 2HDM with flavour blind phases, the so-called ${\rm 2HDM_{\overline{MFV}}}$ \cite{Buras:2010mh,Buras:2010zm} as well. But here the presence of new CP-violating phases 
 changes  the situation radically as now $\vub$ can 
be chosen larger in order to fit $\varepsilon_K$. The enhanced value  of
$\sin 2\beta$ can be suppressed through these new phases. This implies on 
the other hand 
enhancement of the asymmetry $S_{\psi\phi}$ to values $0.15-0.20$  \cite{Buras:2012ts} which 
is $2\sigma$ above the central experimental value from LHCb and could be ruled 
out when the data improve.

Possibly the simplest solution to the problems of various models with MFV 
is to reduce the flavour symmetry $U(3)^3$  to
$U(2)^3$  \cite{Barbieri:2011ci,Barbieri:2011fc,Barbieri:2012uh,Barbieri:2012bh,Crivellin:2011fb,Crivellin:2011sj,Crivellin:2008mq}.
 As pointed out in \cite{Buras:2012sd} in this case NP effects in $\varepsilon_K$ and $B^0_{s,d}-\bar B^0_{s,d}$ are not correlated with each other so that the 
enhancement of $\varepsilon_K$ and suppression of $\Delta M_{s,d}$  
can be achieved if necessary in principle for
the values of  $\vcb$, $F_{B_s} \sqrt{\hat B_{B_s}}$ and $F_{B_d} \sqrt{\hat B_{B_d}}$ in
Table~\ref{tab:input}.

 In particular,
\begin{itemize}
\item
NP effects in $\varepsilon_K$ are of CMFV type and $\varepsilon_K$ 
can only be enhanced but the size of necessary enhancement depends 
on the value of $\vub$ which similar to  ${\rm 2HDM_{\overline{MFV}}}$ 
does not have to be low.
\item
In  $B^0_{s,d}-\bar B^0_{s,d}$ system, the ratio $\Delta M_s/\Delta M_d$ is 
equal to the one in the SM and in good agreement with the data. But 
in view of new CP-violating phases $\varphi_{B_d}$ and $\varphi_{B_s}$ even 
in the presence of only SM operators, $\Delta M_{s,d}$ can be suppressed. 
But the  $U(2)^3$ symmetry implies 
$\varphi_{B_d}=\varphi_{B_s}$ and consequently 
a  triple $S_{\psi K_S}-S_{\psi\phi}-|V_{ub}|$ correlation which constitutes 
and important test of this NP scenario \cite{Buras:2012sd}.
\item
The important advantage  of $U(2)^3$ models over ${\rm 2HDM_{\overline{MFV}}}$ is that in the 
case of $S_{\psi\phi}$ being very small or even having opposite sign to 
SM prediction, this framework can survive with concrete prediction for 
$\vub$.
\end{itemize}

\section{Conclusions}\label{sec:6}
In this paper we have determined $F_{B_s} \sqrt{\hat B_{B_s}}$,  
$F_{B_d} \sqrt{\hat B_{B_d}}$, $\vcb$, $\vub$, $\vtd$ and $\vts$, necessary to fit 
 very precise data on $\varepsilon_K$, $\Delta M_d$, 
$\Delta M_s$ and $S_{\psi K_S}$ 
  in CMFV models as functions of the 
phase $\gamma$ and the sole NP free parameter 
in the $\Delta F=2$ transitions, the value of the box diagram function $S(v)$. 
An important ingredient of this analysis was a very small error on $\hat B_K$ 
from lattice and the fact that  $\hat B_K$ comes within $1-2\%$ from its 
large $N$ value $0.75$.
The results are shown in Table~\ref{tab:CMFVpred}   and   Figs.~\ref{fig:VvsSv}-\ref{fig:VubvsFB}.  The chart showing the execution of our
strategy 
can be found in 
Fig.~\ref{fig:chart}.

Our main messages from this analysis are as follows:
\begin{itemize}
\item
The tension between $\varepsilon_K$ and $\Delta M_{s,d}$ in CMFV models 
accompanied with $|\varepsilon_K|$ being smaller than the data within the 
SM, cannot be removed by varying $S(v)$ when the present input parameters
in Table~\ref{tab:input} are used.
\item
Rather the value of $\vcb$ has to be increased and the values of $F_{B_s} \sqrt{\hat B_{B_s}}$ and $F_{B_d} \sqrt{\hat B_{B_d}}$ decreased relatively to the 
presently quoted lattice values. These enhancements and suppressions are correlated 
with each other and depend on $\gamma$.  The allowed regions in the space 
of these three parameters, the most important results of our paper, 
are  shown in  Figs.~\ref{fig:VcbvsFB}, \ref{fig:VcbvsFBscan} and 
\ref{fig:VcbvsFBscan2}.
\item
 The knowledge of long distance contributions to $\Delta M_K$ accompanied 
by the very precise experimental value of the latter allows a 
significant reduction of the present uncertainty in the value of 
the QCD factor $\eta_{cc}$ under the plausible assumption that $\Delta M_K$ 
in CMFV models is fully dominated by the SM contribution. This  implies
the reduction of the theoretical 
error in $\varepsilon_K$ and in turn the reduction of the error 
in the extraction of the favoured 
value of  $\vcb$ in the CMFV framework. Present estimates of these long 
distance contributions using large $N$ QCD  allow for an optimism 
 but more sophisticated lattice 
calculations are required to fully execute this idea.
\end{itemize}

We have also discussed simplest extensions of the SM which could in principle 
offer a better description of the data in case CMFV would fail to do so. 
Models with $U(2)^3$ flavour symmetry appear to us as most efficient in 
this respect while being still very simple. In order to see
whether the CMFV framework will survive final tests from $\Delta F=2$ 
transitions further progress from lattice calculations and experimental 
measurements is required. Our wish list includes:
\begin{itemize}
\item
In particular improved lattice calculations of $F_{B_s}\sqrt{\hat B_{B_s}}$, $F_{B_d} \sqrt{\hat B_{B_d}}$ and $\xi$ but also of $\hat B_K$, $\hat B_{B_s}$ and $\hat B_{B_s}$.
\item
Calculation of long distance contribution to $\Delta M_K$ in order to reduce 
the error on $\eta_{cc}$ as proposed by us.
\item
Improved experimental data on $\vcb$, $\vub$, 
$\gamma$,  $S_{\psi K_S}$ and $S_{\psi\phi}$. In particular a measurement 
of $S_{\psi\phi}$ significantly different from $0.04$ would signal the 
presence of new phases beyond the CMFV framework.
\end{itemize}

The correlations identified in this paper will allow to 
monitor the future developments, likely to indicate that new sources of flavour and CP-violation  beyond the CMFV framework are present in nature. In this 
context our main message is the following one. While until now the search 
for NP through rare decays did not bring any convincing signs of its presence, 
it could turn out soon that $\Delta F=2$ transitions combined with 
the progress made by lattice community will herald the appearance of particular 
type of NP. This would not only be exciting news but would also give some 
 directions for searching for this NP in rare decays and even high-energy 
processes.

{\bf Acknowledgements}\\
We thank Jean-Marc G\'erard for many illuminating comments and Chris Sachrajda 
for useful information on the progress in lattice calculations.
This research was fully financed and done in the context of the ERC Advanced Grant project ``FLAVOUR'' (267104).

\bibliographystyle{JHEP}
\bibliography{allrefs}

\providecommand{\href}[2]{#2}\begingroup\raggedright\begin{thebibliography}{10}

\bibitem{Buras:2000dm}
A.~J. Buras, P.~Gambino, M.~Gorbahn, S.~Jager, and L.~Silvestrini, {\it
  Universal unitarity triangle and physics beyond the standard model},  {\em
  Phys. Lett.} {\bf B500} (2001) 161--167,
  [\href{http://xxx.lanl.gov/abs/hep-ph/0007085}{{\tt hep-ph/0007085}}].

\bibitem{Buras:2003jf}
A.~J. Buras, {\it Minimal flavor violation},  {\em Acta Phys. Polon.} {\bf B34}
  (2003) 5615--5668, [\href{http://xxx.lanl.gov/abs/hep-ph/0310208}{{\tt
  hep-ph/0310208}}].

\bibitem{Blanke:2006ig}
M.~Blanke, A.~J. Buras, D.~Guadagnoli, and C.~Tarantino, {\it {Minimal Flavour
  Violation Waiting for Precise Measurements of $\Delta M_s$, $S_{\psi \phi}$,
  $A^s_\text{SL}$, $|V_{ub}|$, $\gamma$ and $B^0_{s,d} \to \mu^+ \mu^-$}},
  {\em JHEP} {\bf 10} (2006) 003,
  [\href{http://xxx.lanl.gov/abs/hep-ph/0604057}{{\tt hep-ph/0604057}}].

\bibitem{Blanke:2006yh}
M.~Blanke and A.~J. Buras, {\it {Lower bounds on $\Delta M_{s,d}$ from
  constrained minimal flavour violation}},  {\em JHEP} {\bf 0705} (2007) 061,
  [\href{http://xxx.lanl.gov/abs/hep-ph/0610037}{{\tt hep-ph/0610037}}].

\bibitem{Lunghi:2008aa}
E.~Lunghi and A.~Soni, {\it {Possible Indications of New Physics in
  $B_d$-mixing and in $\sin(2 \beta)$ Determinations}},  {\em Phys. Lett.} {\bf
  B666} (2008) 162--165, [\href{http://xxx.lanl.gov/abs/0803.4340}{{\tt
  arXiv:0803.4340}}].

\bibitem{Buras:2008nn}
A.~J. Buras and D.~Guadagnoli, {\it {Correlations among new CP violating
  effects in $\Delta F = 2$ observables}},  {\em Phys. Rev.} {\bf D78} (2008)
  033005, [\href{http://xxx.lanl.gov/abs/0805.3887}{{\tt arXiv:0805.3887}}].

\bibitem{Bona:2009cj}
{\bf UTfit Collaboration} Collaboration, M.~Bona {\em et.~al.}, {\it {An
  Improved Standard Model Prediction Of $BR(B\to \tau \nu)$ And Its
  Implications For New Physics}},  {\em Phys.Lett.} {\bf B687} (2010) 61--69,
  [\href{http://xxx.lanl.gov/abs/0908.3470}{{\tt arXiv:0908.3470}}].

\bibitem{Lenz:2010gu}
A.~Lenz, U.~Nierste, J.~Charles, S.~Descotes-Genon, A.~Jantsch, {\em et.~al.},
  {\it {Anatomy of New Physics in $B - \bar{B}$ mixing}},  {\em Phys.Rev.} {\bf
  D83} (2011) 036004, [\href{http://xxx.lanl.gov/abs/1008.1593}{{\tt
  arXiv:1008.1593}}].

\bibitem{Lunghi:2010gv}
E.~Lunghi and A.~Soni, {\it {Possible evidence for the breakdown of the
  CKM-paradigm of CP-violation}},  {\em Phys.Lett.} {\bf B697} (2011) 323--328,
  [\href{http://xxx.lanl.gov/abs/1010.6069}{{\tt arXiv:1010.6069}}].

\bibitem{Buras:2011wi}
A.~J. Buras, M.~V. Carlucci, L.~Merlo, and E.~Stamou, {\it {Phenomenology of a
  Gauged $SU(3)^3$ Flavour Model}},
  \href{http://xxx.lanl.gov/abs/1112.4477}{{\tt arXiv:1112.4477}}.

\bibitem{Buras:2012ts}
A.~J. Buras and J.~Girrbach, {\it {BSM models facing the recent LHCb data: A
  First look}},  {\em Acta Phys.Polon.} {\bf B43} (2012) 1427,
  [\href{http://xxx.lanl.gov/abs/1204.5064}{{\tt arXiv:1204.5064}}].

\bibitem{Davies:2012qf}
C.~Davies, {\it {Standard Model Heavy Flavor physics on the Lattice}},  {\em
  PoS} {\bf LATTICE2011} (2011) 019,
  [\href{http://xxx.lanl.gov/abs/1203.3862}{{\tt arXiv:1203.3862}}].

\bibitem{Gamiz:2013waa}
E.~G\'amiz, {\it {Flavour physics from lattice QCD}},
  \href{http://xxx.lanl.gov/abs/1303.3971}{{\tt arXiv:1303.3971}}.

\bibitem{Aoki:2010pe}
Y.~Aoki, R.~Arthur, T.~Blum, P.~Boyle, D.~Brommel, {\em et.~al.}, {\it
  {Continuum Limit of $B_K$ from 2+1 Flavor Domain Wall QCD}},  {\em Phys.Rev.}
  {\bf D84} (2011) 014503, [\href{http://xxx.lanl.gov/abs/1012.4178}{{\tt
  arXiv:1012.4178}}].

\bibitem{Bae:2010ki}
T.~Bae, Y.-C. Jang, C.~Jung, H.-J. Kim, J.~Kim, {\em et.~al.}, {\it {$B_K$
  using HYP-smeared staggered fermions in $N_f=2+1$ unquenched QCD}},  {\em
  Phys.Rev.} {\bf D82} (2010) 114509,
  [\href{http://xxx.lanl.gov/abs/1008.5179}{{\tt arXiv:1008.5179}}].

\bibitem{Constantinou:2010qv}
{\bf ETM Collaboration} Collaboration, M.~Constantinou {\em et.~al.}, {\it
  {$B_K$-parameter from $N_f$ = 2 twisted mass lattice QCD}},  {\em Phys.Rev.}
  {\bf D83} (2011) 014505, [\href{http://xxx.lanl.gov/abs/1009.5606}{{\tt
  arXiv:1009.5606}}].

\bibitem{Colangelo:2010et}
G.~Colangelo, S.~Durr, A.~Juttner, L.~Lellouch, H.~Leutwyler, {\em et.~al.},
  {\it {Review of lattice results concerning low energy particle physics}},
  {\em Eur.Phys.J.} {\bf C71} (2011) 1695,
  [\href{http://xxx.lanl.gov/abs/1011.4408}{{\tt arXiv:1011.4408}}].

\bibitem{Bailey:2012bh}
J.~A. Bailey, T.~Bae, Y.-C. Jang, H.~Jeong, C.~Jung, {\em et.~al.}, {\it
  {Beyond the Standard Model corrections to $K^0-\bar{K}^0$ mixing}},  {\em
  PoS} {\bf LATTICE2012} (2012) 107,
  [\href{http://xxx.lanl.gov/abs/1211.1101}{{\tt arXiv:1211.1101}}].

\bibitem{Durr:2011ap}
S.~Durr, Z.~Fodor, C.~Hoelbling, S.~Katz, S.~Krieg, {\em et.~al.}, {\it
  {Precision computation of the kaon bag parameter}},  {\em Phys.Lett.} {\bf
  B705} (2011) 477--481, [\href{http://xxx.lanl.gov/abs/1106.3230}{{\tt
  arXiv:1106.3230}}].

\bibitem{Laiho:2009eu}
J.~Laiho, E.~Lunghi, and R.~S. Van~de Water, {\it {Lattice QCD inputs to the
  CKM unitarity triangle analysis}},  {\em Phys. Rev.} {\bf D81} (2010) 034503,
  [\href{http://xxx.lanl.gov/abs/0910.2928}{{\tt arXiv:0910.2928}}]. Updates
  available on {\tt http://latticeaverages.org/}.

\bibitem{Gaiser:1980gx}
B.~D. Gaiser, T.~Tsao, and M.~B. Wise, {\it {Parameters of the six quark
  model}},  {\em Annals Phys.} {\bf 132} (1981) 66.

\bibitem{Buras:1985yx}
A.~J. Buras and J.-M. G\'erard, {\it {$1/N$ Expansion for Kaons}},  {\em
  Nucl.Phys.} {\bf B264} (1986) 371.

\bibitem{Bardeen:1987vg}
W.~A. Bardeen, A.~J. Buras, and J.-M. G\'erard, {\it {The B Parameter Beyond
  the Leading Order of 1/N Expansion}},  {\em Phys.Lett.} {\bf B211} (1988)
  343.

\bibitem{Gerard:2010jt}
J.-M. G\'erard, {\it {An upper bound on the Kaon B-parameter and ${\rm
  Re}(\epsilon_K)$}},  {\em JHEP} {\bf 1102} (2011) 075,
  [\href{http://xxx.lanl.gov/abs/1012.2026}{{\tt arXiv:1012.2026}}].

\bibitem{Bona:2007vi}
{\bf UTfit Collaboration} Collaboration, M.~Bona {\em et.~al.}, {\it
  {Model-independent constraints on $\Delta$ F=2 operators and the scale of new
  physics}},  {\em JHEP} {\bf 0803} (2008) 049,
  [\href{http://xxx.lanl.gov/abs/0707.0636}{{\tt arXiv:0707.0636}}]. {Updates
  available on \texttt{http://www.utfit.org}.}

\bibitem{Buras:2012sd}
A.~J. Buras and J.~Girrbach, {\it {On the Correlations between Flavour
  Observables in Minimal $U(2)^3$ Models}},  {\em JHEP} {\bf 1301} (2013) 007,
  [\href{http://xxx.lanl.gov/abs/1206.3878}{{\tt arXiv:1206.3878}}].

\bibitem{Buras:2012dp}
A.~J. Buras, F.~De~Fazio, J.~Girrbach, and M.~V. Carlucci, {\it {The Anatomy of
  Quark Flavour Observables in 331 Models in the Flavour Precision Era}},  {\em
  JHEP} {\bf 1302} (2013) 023, [\href{http://xxx.lanl.gov/abs/1211.1237}{{\tt
  arXiv:1211.1237}}].

\bibitem{Buras:2012jb}
A.~J. Buras, F.~De~Fazio, and J.~Girrbach, {\it {The Anatomy of Z' and Z with
  Flavour Changing Neutral Currents in the Flavour Precision Era}},  {\em JHEP}
  {\bf 1302} (2013) 116, [\href{http://xxx.lanl.gov/abs/1211.1896}{{\tt
  arXiv:1211.1896}}].

\bibitem{Buras:2013td}
A.~J. Buras, J.~Girrbach, and R.~Ziegler, {\it {Particle-Antiparticle Mixing,
  CP Violation and Rare K and B Decays in a Minimal Theory of Fermion Masses}},
   \href{http://xxx.lanl.gov/abs/1301.5498}{{\tt arXiv:1301.5498}}.

\bibitem{Buras:2013uqa}
A.~J. Buras, R.~Fleischer, J.~Girrbach, and R.~Knegjens, {\it {Probing New
  Physics with the $B_s\to\mu^+\mu^-$ Time-Dependent Rate}},
  \href{http://xxx.lanl.gov/abs/1303.3820}{{\tt arXiv:1303.3820}}.

\bibitem{Buras:2013rqa}
A.~J. Buras, F.~De~Fazio, J.~Girrbach, R.~Knegjens, and M.~Nagai, {\it {The
  Anatomy of Neutral Scalars with FCNCs in the Flavour Precision Era}},
  \href{http://xxx.lanl.gov/abs/1303.3723}{{\tt arXiv:1303.3723}}.

\bibitem{Kettell:2002ep}
S.~H. Kettell, L.~Landsberg, and H.~H. Nguyen, {\it {Alternative technique for
  standard model estimation of the rare kaon decay branchings BR($K \to \pi \nu
  \bar{\nu}$) (SM)}},  {\em Phys.Atom.Nucl.} {\bf 67} (2004) 1398--1407,
  [\href{http://xxx.lanl.gov/abs/hep-ph/0212321}{{\tt hep-ph/0212321}}].

\bibitem{Buras:2002yj}
A.~J. Buras, F.~Parodi, and A.~Stocchi, {\it {The CKM matrix and the unitarity
  triangle: Another look}},  {\em JHEP} {\bf 0301} (2003) 029,
  [\href{http://xxx.lanl.gov/abs/hep-ph/0207101}{{\tt hep-ph/0207101}}].

\bibitem{Buras:2010pza}
A.~J. Buras, D.~Guadagnoli, and G.~Isidori, {\it {On $\epsilon_K$ beyond lowest
  order in the Operator Product Expansion}},  {\em Phys.Lett.} {\bf B688}
  (2010) 309--313, [\href{http://xxx.lanl.gov/abs/1002.3612}{{\tt
  arXiv:1002.3612}}].

\bibitem{Blanke:2011ry}
M.~Blanke, A.~J. Buras, K.~Gemmler, and T.~Heidsieck, {\it {$\Delta F = 2$
  observables and $B\to X_q\gamma$ in the Left-Right Asymmetric Model: Higgs
  particles striking back}},  {\em JHEP} {\bf 1203} (2012) 024,
  [\href{http://xxx.lanl.gov/abs/1111.5014}{{\tt arXiv:1111.5014}}].

\bibitem{Dowdall:2013tga}
R.~Dowdall, C.~Davies, R.~Horgan, C.~Monahan, and J.~Shigemitsu, {\it {B-meson
  decay constants from improved lattice NRQCD and physical u, d, s and c sea
  quarks}},  \href{http://xxx.lanl.gov/abs/1302.2644}{{\tt arXiv:1302.2644}}.

\bibitem{Nakamura:2010zzi}
{\bf Particle Data Group} Collaboration, K.~Nakamura {\em et.~al.}, {\it
  {Review of particle physics}},  {\em J.Phys.G} {\bf G37} (2010) 075021.

\bibitem{Beringer:1900zz}
{\bf Particle Data Group} Collaboration, J.~Beringer {\em et.~al.}, {\it
  {Review of Particle Physics (RPP)}},  {\em Phys.Rev.} {\bf D86} (2012)
  010001.

\bibitem{Chetyrkin:2009fv}
K.~Chetyrkin, J.~Kuhn, A.~Maier, P.~Maierhofer, P.~Marquard, {\em et.~al.},
  {\it {Charm and Bottom Quark Masses: An Update}},  {\em Phys.Rev.} {\bf D80}
  (2009) 074010, [\href{http://xxx.lanl.gov/abs/0907.2110}{{\tt
  arXiv:0907.2110}}].

\bibitem{Allison:2008xk}
{\bf HPQCD Collaboration} Collaboration, I.~Allison {\em et.~al.}, {\it
  {High-Precision Charm-Quark Mass from Current-Current Correlators in Lattice
  and Continuum QCD}},  {\em Phys.Rev.} {\bf D78} (2008) 054513,
  [\href{http://xxx.lanl.gov/abs/0805.2999}{{\tt arXiv:0805.2999}}].

\bibitem{Buras:1990fn}
A.~J. Buras, M.~Jamin, and P.~H. Weisz, {\it {Leading and next-to-leading QCD
  corrections to $\varepsilon$ parameter and $B^0-\bar{B}^0$ mixing in the
  presence of a heavy top quark}},  {\em Nucl. Phys.} {\bf B347} (1990)
  491--536.

\bibitem{Urban:1997gw}
J.~Urban, F.~Krauss, U.~Jentschura, and G.~Soff, {\it {Next-to-leading order
  QCD corrections for the $B^0 - \bar B^0$ mixing with an extended Higgs
  sector}},  {\em Nucl. Phys.} {\bf B523} (1998) 40--58,
  [\href{http://xxx.lanl.gov/abs/hep-ph/9710245}{{\tt hep-ph/9710245}}].

\bibitem{Aaltonen:2012ra}
{\bf CDF Collaboration, D0 Collaboration} Collaboration, T.~Aaltonen {\em
  et.~al.}, {\it {Combination of the top-quark mass measurements from the
  Tevatron collider}},  {\em Phys.Rev.} {\bf D86} (2012) 092003,
  [\href{http://xxx.lanl.gov/abs/1207.1069}{{\tt arXiv:1207.1069}}].

\bibitem{Amhis:2012bh}
{\bf Heavy Flavor Averaging Group} Collaboration, Y.~Amhis {\em et.~al.}, {\it
  {Averages of B-Hadron, C-Hadron, and tau-lepton properties as of early
  2012}},  \href{http://xxx.lanl.gov/abs/1207.1158}{{\tt arXiv:1207.1158}}.

\bibitem{Aaij:2013oba}
{\bf LHCb collaboration} Collaboration, R.~Aaij {\em et.~al.}, {\it
  {Measurement of $CP$ violation and the $B_s^0$ meson decay width difference
  with $B_s^0 \to J/\psi K^+K^-$ and $B_s^0\to J/\psi\pi^+\pi^-$ decays}},
  \href{http://xxx.lanl.gov/abs/1304.2600}{{\tt arXiv:1304.2600}}.

\bibitem{Raven:2012fb}
{\bf LHCb Collaboration} Collaboration, G.~Raven, {\it {Measurement of the CP
  violation phase $\phi_s$ in the $B_s$ system at LHCb}},
  \href{http://xxx.lanl.gov/abs/1212.4140}{{\tt arXiv:1212.4140}}.

\bibitem{Brod:2011ty}
J.~Brod and M.~Gorbahn, {\it {Next-to-Next-to-Leading-Order Charm-Quark
  Contribution to the CP Violation Parameter $\varepsilon_K$ and $\Delta
  M_K$}},  {\em Phys.Rev.Lett.} {\bf 108} (2012) 121801,
  [\href{http://xxx.lanl.gov/abs/1108.2036}{{\tt arXiv:1108.2036}}].

\bibitem{Brod:2010mj}
J.~Brod and M.~Gorbahn, {\it {$\epsilon_K$ at Next-to-Next-to-Leading Order:
  The Charm-Top-Quark Contribution}},  {\em Phys.Rev.} {\bf D82} (2010) 094026,
  [\href{http://xxx.lanl.gov/abs/1007.0684}{{\tt arXiv:1007.0684}}].

\bibitem{Tarantino:2012mq}
C.~Tarantino, {\it {Flavor Lattice QCD in the Precision Era}},
  \href{http://xxx.lanl.gov/abs/1210.0474}{{\tt arXiv:1210.0474}}.

\bibitem{Fleischer:2010ib}
R.~Fleischer and R.~Knegjens, {\it {In Pursuit of New Physics with $B^0_s\to
  K^+K^-$}},  {\em Eur.Phys.J.} {\bf C71} (2011) 1532,
  [\href{http://xxx.lanl.gov/abs/1011.1096}{{\tt arXiv:1011.1096}}].

\bibitem{'tHooft:1973jz}
G.~'t~Hooft, {\it {A Planar Diagram Theory for Strong Interactions}},  {\em
  Nucl.Phys.} {\bf B72} (1974) 461.

\bibitem{'tHooft:1974hx}
G.~'t~Hooft, {\it {A Two-Dimensional Model for Mesons}},  {\em Nucl.Phys.} {\bf
  B75} (1974) 461.

\bibitem{Witten:1979kh}
E.~Witten, {\it {Baryons in the 1/n Expansion}},  {\em Nucl.Phys.} {\bf B160}
  (1979) 57.

\bibitem{Bardeen:1986vp}
W.~A. Bardeen, A.~J. Buras, and J.-M. G\'erard, {\it {The $\Delta I = 1/2$ Rule
  in the Large $N$ Limit}},  {\em Phys.Lett.} {\bf B180} (1986) 133.

\bibitem{Bardeen:1986uz}
W.~A. Bardeen, A.~J. Buras, and J.-M. G\'erard, {\it {The $K\to\pi \pi$ Decays
  in the Large N Limit: Quark Evolution}},  {\em Nucl.Phys.} {\bf B293} (1987)
  787.

\bibitem{Bardeen:1986vz}
W.~A. Bardeen, A.~J. Buras, and J.-M. G\'erard, {\it {A Consistent Analysis of
  the $\Delta I = 1/2$ Rule for K Decays}},  {\em Phys.Lett.} {\bf B192} (1987)
  138.

\bibitem{Blum:2012uk}
T.~Blum, P.~Boyle, N.~Christ, N.~Garron, E.~Goode, {\em et.~al.}, {\it {Lattice
  determination of the $K \to (\pi\pi)_{I=2}$ Decay Amplitude $A_2$}},
  \href{http://xxx.lanl.gov/abs/1206.5142}{{\tt arXiv:1206.5142}}.

\bibitem{Boyle:2012ys}
{\bf RBC Collaboration, UKQCD Collaboration} Collaboration, P.~Boyle {\em
  et.~al.}, {\it {Emerging understanding of the $\Delta I = 1/2$ Rule from
  Lattice QCD}},  \href{http://xxx.lanl.gov/abs/1212.1474}{{\tt
  arXiv:1212.1474}}.

\bibitem{Bijnens:1990mz}
J.~Bijnens, J.-M. G\'erard, and G.~Klein, {\it {The $K_L - K_S$ mass
  difference}},  {\em Phys.Lett.} {\bf B257} (1991) 191--195.

\bibitem{Gerard:1990dx}
J.-M. G\'erard, {\it {Electroweak interactions of hadrons}},  {\em Acta
  Phys.Polon.} {\bf B21} (1990) 257--305.

\bibitem{Donoghue:1983hi}
J.~F. Donoghue, E.~Golowich, and B.~R. Holstein, {\it {Long Distance chiral
  contributions to $K_L-K_S$ mass difference}},  {\em Phys.Lett.} {\bf B135}
  (1984) 481.

\bibitem{Gerard:2005yk}
J.-M. G\'erard, C.~Smith, and S.~Trine, {\it {Radiative kaon decays and the
  penguin contribution to the $\Delta I = 1/2$ rule}},  {\em Nucl.Phys.} {\bf
  B730} (2005) 1--36, [\href{http://xxx.lanl.gov/abs/hep-ph/0508189}{{\tt
  hep-ph/0508189}}].

\bibitem{Christ:2012se}
N.~Christ, T.~Izubuchi, C.~Sachrajda, A.~Soni, and J.~Yu, {\it {Long distance
  contribution to the $K_L-K_S$ mass difference}},
  \href{http://xxx.lanl.gov/abs/1212.5931}{{\tt arXiv:1212.5931}}.

\bibitem{Chivukula:1987py}
R.~S. Chivukula and H.~Georgi, {\it Composite technicolor standard model},
  {\em Phys. Lett.} {\bf B188} (1987) 99.

\bibitem{Hall:1990ac}
L.~J. Hall and L.~Randall, {\it Weak scale effective supersymmetry},  {\em
  Phys. Rev. Lett.} {\bf 65} (1990) 2939--2942.

\bibitem{D'Ambrosio:2002ex}
G.~D'Ambrosio, G.~F. Giudice, G.~Isidori, and A.~Strumia, {\it {Minimal flavour
  violation: An effective field theory approach}},  {\em Nucl. Phys.} {\bf
  B645} (2002) 155--187, [\href{http://xxx.lanl.gov/abs/hep-ph/0207036}{{\tt
  hep-ph/0207036}}].

\bibitem{Buras:2002wq}
A.~J. Buras, P.~H. Chankowski, J.~Rosiek, and L.~Slawianowska, {\it Correlation
  between $\delta m_{s}$ and $b_{d,s}^0 \to\mu^+ \mu^-$ in supersymmetry at
  large $\tan\beta$},  {\em Phys. Lett.} {\bf B546} (2002) 96--107,
  [\href{http://xxx.lanl.gov/abs/hep-ph/0207241}{{\tt hep-ph/0207241}}].

\bibitem{Buras:2010mh}
A.~J. Buras, M.~V. Carlucci, S.~Gori, and G.~Isidori, {\it {Higgs-mediated
  FCNCs: Natural Flavour Conservation vs. Minimal Flavour Violation}},  {\em
  JHEP} {\bf 1010} (2010) 009, [\href{http://xxx.lanl.gov/abs/1005.5310}{{\tt
  arXiv:1005.5310}}].

\bibitem{Buras:2010zm}
A.~J. Buras, G.~Isidori, and P.~Paradisi, {\it {EDMs versus CPV in $B_{s,d}$
  mixing in two Higgs doublet models with MFV}},  {\em Phys.Lett.} {\bf B694}
  (2011) 402--409, [\href{http://xxx.lanl.gov/abs/1007.5291}{{\tt
  arXiv:1007.5291}}].

\bibitem{Barbieri:2011ci}
R.~Barbieri, G.~Isidori, J.~Jones-Perez, P.~Lodone, and D.~M. Straub, {\it
  {U(2) and Minimal Flavour Violation in Supersymmetry}},  {\em Eur.Phys.J.}
  {\bf C71} (2011) 1725, [\href{http://xxx.lanl.gov/abs/1105.2296}{{\tt
  arXiv:1105.2296}}].

\bibitem{Barbieri:2011fc}
R.~Barbieri, P.~Campli, G.~Isidori, F.~Sala, and D.~M. Straub, {\it {B-decay
  CP-asymmetries in SUSY with a $U(2)^3$ flavour symmetry}},  {\em Eur.Phys.J.}
  {\bf C71} (2011) 1812, [\href{http://xxx.lanl.gov/abs/1108.5125}{{\tt
  arXiv:1108.5125}}].

\bibitem{Barbieri:2012uh}
R.~Barbieri, D.~Buttazzo, F.~Sala, and D.~M. Straub, {\it {Flavour physics from
  an approximate $U(2)^3$ symmetry}},  {\em JHEP} {\bf 1207} (2012) 181,
  [\href{http://xxx.lanl.gov/abs/1203.4218}{{\tt arXiv:1203.4218}}].

\bibitem{Barbieri:2012bh}
R.~Barbieri, D.~Buttazzo, F.~Sala, and D.~M. Straub, {\it {Less Minimal Flavour
  Violation}},  {\em JHEP} {\bf 1210} (2012) 040,
  [\href{http://xxx.lanl.gov/abs/1206.1327}{{\tt arXiv:1206.1327}}].

\bibitem{Crivellin:2011fb}
A.~Crivellin, L.~Hofer, and U.~Nierste, {\it {The MSSM with a Softly Broken
  $U(2)^3$ Flavor Symmetry}},  {\em PoS} {\bf EPS-HEP2011} (2011) 145,
  [\href{http://xxx.lanl.gov/abs/1111.0246}{{\tt arXiv:1111.0246}}].

\bibitem{Crivellin:2011sj}
A.~Crivellin, L.~Hofer, U.~Nierste, and D.~Scherer, {\it {Phenomenological
  consequences of radiative flavor violation in the MSSM}},  {\em Phys.Rev.}
  {\bf D84} (2011) 035030, [\href{http://xxx.lanl.gov/abs/1105.2818}{{\tt
  arXiv:1105.2818}}].

\bibitem{Crivellin:2008mq}
A.~Crivellin and U.~Nierste, {\it {Supersymmetric renormalisation of the CKM
  matrix and new constraints on the squark mass matrices}},  {\em Phys.Rev.}
  {\bf D79} (2009) 035018, [\href{http://xxx.lanl.gov/abs/0810.1613}{{\tt
  arXiv:0810.1613}}].

\end{thebibliography}\endgroup
\end{document}